\documentclass[11pt]{article}

\usepackage{amsfonts,graphicx,amsmath,amssymb,amsthm,geometry,bbm,hyperref,nicefrac,enumerate,comment,fontawesome,mathtools,booktabs,siunitx,dirtytalk,mathrsfs}

\usepackage{xcolor}
\DeclareMathOperator*{\argmax}{arg\,max}
\DeclareMathOperator*{\argmin}{arg\,min}

\newcommand{\I}{\mathbb{I}}
\newcommand{\R}{\mathbb{R}}
\newcommand{\N}{\mathbb{N}}
\newcommand{\E}{\mathbb{E}}

\renewcommand{\P}{\mathbb{P}}
\newcommand{\Q}{\mathbb{Q}}

\newcommand{\T}{\mathcal{T}}

\newcommand{\F}{\mathcal{F}}

\newcommand{\Ft}{\mathcal{F}_t}
\newcommand{\Ht}{\mathcal{H}_t}
\newcommand{\e}{\text{e}}
\renewcommand{\d}{\text{d}}
\newcommand{\D}{\mathcal{D}}
\usepackage{array}
\newcolumntype{P}[1]{>{\centering\arraybackslash}p{#1}}

\usepackage{geometry}
\newtheorem{remark}{Remark}[section]

\newtheorem{example}{Example}[section]

\theoremstyle{definition}

\newcommand{\footremember}[2]{%
    \footnote{#2}
    \newcounter{#1}
    \setcounter{#1}{\value{footnote}}%
}
\newcommand{\footrecall}[1]{%
    \footnotemark[\value{#1}]%
} 

\begin{document}

\title{Deep learning for CVA computations of large portfolios of financial derivatives}

\author{%
  Kristoffer Andersson\footremember{alley}{Research Group of Scientific Computing, Centrum Wiskunde \& Informatica} 
  \and Cornelis W. Oosterlee\footrecall{alley} \footremember{trailer}{Delft Institute of Applied Mathematics (DIAM), Delft University of Technology}
 }

\maketitle
\begin{abstract}

In this paper, we propose a neural network-based method for CVA computations of a portfolio of derivatives. In particular, we focus on portfolios consisting of a combination of derivatives, with and without true optionality, \textit{e.g.,} a portfolio of a mix of European- and Bermudan-type derivatives. CVA is computed, with and without netting, for different levels of WWR and for different levels of credit quality of the counterparty. We show that the CVA is overestimated with up to 25\% by using the standard procedure of not adjusting the exercise strategy for the default-risk of the counterparty. For the Expected Shortfall of the CVA dynamics, the overestimation was found to be more than 100\% in some non-extreme cases.

\end{abstract}
\tableofcontents

\section{Introduction}
In this paper, we consider a set of financial contracts, which we refer to as the \textit{portfolio of derivatives}, or just the \textit{portfolio}, written between two parties. The first party is referred to as the \textit{bank} and is considered to be default-free. The second party, which may default, is referred to as the \textit{counterparty}. We take the perspective of the default-free bank in order to investigate some of the risks associated with a defaultable counterparty. It is straight-forward to extend the methodologies used in this paper to a defaultable bank as well as to multiple counterparties. 

\subsection{Risk-free valuation}
We consider the problem of finding the value of a portfolio of derivatives with early-exercise features. In particular, we are focusing on portfolios with multiple derivatives with true optionality, \textit{e.g.,} American or Bermudan derivatives. We construct a portfolio of $J$ derivatives, where the individual derivatives depend on $d_1,d_2,\ldots,d_J$ risk factors. This means that we could face high-dimensionality in two ways:
\begin{enumerate}
    \item Derivative $j$ could depend on a large number of risk factors, \textit{i.e.,} $d_j$ could be large;
    \item We could have many derivatives in the portfolio, \textit{i.e.,} $J$ could be large.
\end{enumerate}
In \cite{DOS}, a neural network-based method for valuation of a single Bermudan derivative was proposed and proved to be highly accurate for derivatives with up to $100$ risk factors. Later, the algorithm was extended in \cite{DOSexposure} to also include pathwise valuations of the derivative (in contrast to only finding the value at the initial time). In this paper, we extend \cite{DOS} and \cite{DOSexposure} to the portfolio case, \textit{i.e.,} finding the value of a large portfolio of, possibly high-dimensional, derivatives with true optionality, without having to compute the value of each individual derivative.

In a traditional setting, the so-called continuation value is computed, and subsequently, the value of the derivative is given by the maximum of the continuation value and the immediate pay-off. For a single derivative, this is straight-forward. For instance, the continuation value can be computed by solving an associated PDE, which is done in \textit{e.g.,} \cite{Forsyth}, \cite{Reisinger}, \cite{Vasquez}, \cite{Hout} and \cite{Hout_book}, or the continuation value can be approximated by a Fourier transform methodology, which is done in \text{e.g.,} \cite{Fang}, \cite{Zy} and \cite{Fang2}. Furthermore, classical tree-based methods such as \cite{tree_broadie}, \cite{Rubinstein} and \cite{Jack}, can be used. These types of methods are, in general, highly accurate but they suffer severely from the curse of dimensionality, meaning that they are computationally feasible only in low dimensions (say up to $4$ risk factors), see \cite{COD}. In higher dimensions, Monte-Carlo-based methods are often used, see \textit{e.g.,} \cite{Binomial_price}, \cite{LSM}, \cite{BG}, \cite{maxcall} and \cite{SGBM}. Monte-Carlo-based methods can generate highly accurate derivative values at the initial time, but often less accurate values between the initial time and maturity of the contract.

In contrast to the single derivative case, it is not enough to know the continuation value of a portfolio (with more than one derivative) in order to decide optimally which derivatives should be exercised. Therefore, it is common to do the valuation at the level of each derivative, and then add the individual values of each derivative to obtain the portfolio value. This becomes cumbersome for large portfolios. As mentioned above, the methodology used in this paper, generalizes \cite{DOS} and \cite{DOSexposure}, in which the optimal exercise policy is approximated by maximizing expected discounted cash-flows, \textit{i.e.,} the continuation value is not computed. By not relying on computations of the continuation value, the algorithm is able to compute the portfolio value without having to compute the individual values for each derivative.  

\subsection{Risky valuation and CVA}
The Credit Valuation Adjustment (CVA) is the difference between the risk-free portfolio value and the risky portfolio value, where the risky portfolio value is defined as the portfolio value when taking default risk of the counterparty into account. While there is no ambiguity of the risk-free portfolio value, it is not completely clear how the risky portfolio value should be computed. The question is whether the exercise policy should be adjusted for the fact that the counterparty may default. For instance, if the counterparty ends up in financial distress, it is reasonable to assume that the bank (which in this paper is assumed to be the risk-free party) would be more willing to exercise the callable derivatives, in order to lower its exposure to the counterparty. Even though it seems common to ignore the effect of a defaultable counterparty when computing risky derivative values, it has been discussed in the literature, see \textit{e.g.,} \cite{CVA_BER0}, \cite{CVA_BER1}, \cite{CVA_BER2} and \cite{CVA_BER3}. In the case of a single derivative \cite{CVA_BER0} states that the exercise region for a risk-free derivative is always a subset of the exercise region for a risky counterpart. However, in the case of a portfolio, the situation is more complex, and depends on contractual details such as the close-out and netting agreements. One consequence is that, in the presence of a netting agreement, the exercise decisions can no longer be made individually. To explain this, we give a simple example. \begin{example}Assume that we have a portfolio, consisting of three derivatives, one European future and two American options. All contracts are initialized at time 0, mature at time $T$ and depend on the same risk-factor $(X_t)_{t\in[0,T]}$. Assume that, at time $t\in(0,T)$, and given $X_t=x$, the intrinsic values are \begin{equation*}
    V^\mathrm{future}(t,x)=-10,\quad V^\mathrm{Am}_1(t,x)=10,\quad V^\mathrm{Am}_2(t,x)=10,
\end{equation*}
and the immediate pay-off for the American options satisfy
\begin{equation*}
    g^\mathrm{Am}_1(t,x)<10,\quad g^\mathrm{Am}_2(t,x)<10.
\end{equation*}
In a risk free environment (no-defaultable counterparty), it is sub-optimal to exercise the American options. However, in case of a defaultable counterparty, the situation is less trivial. In Table \ref{exposure_table}, the exposure to the counterparty, given different exercise decisions at $t$, is given with and without a netting agreement.
\begin{table}[htp]\centering
\begin{tabular}{|l|l|l|}
\hline
 & \text{Without netting} & \text{With netting}  \\ \hline
     \text{Exposure - no exercise}     &   20        &   10   \\ \hline
     \text{Exposure - exercise one of the American options} &  10 & 0  \\ \hline
      \text{Exposure - exercise both American options}    &  0 &  0 \\ \hline
\end{tabular}
\caption{Exposure given different exercise decisions at $t$, with and without a netting agreement.}\label{exposure_table}
\end{table}
If the counterparty is in severe financial distress, then it is likely optimal for the bank to exercise both American options in the case of no netting agreement, and one of them in the case of a netting agreement. From this simple example, two things become clear; 1) The exercise decisions for the American options are affected not only by a risky counterparty, but also by whether or not a netting agreement exists. 2) in the presence of netting, exercise decisions cannot be made for one derivative in isolation, but only for all the American options simultaneously.
\end{example}

In general, for a risky portfolio, it is not possible to describe the value of a single derivative, but only the value of the entire portfolio. This is an interesting problem since almost all existing algorithms rely on exercise decisions made in isolation and risky derivative values that can be added up to obtain the risky portfolio value. 

If this is not taken into account we would obtain a biased low valuation for the risky portfolio by using a sub-optimal exercise strategy. Since the CVA is the difference between the risk-free and risky portfolio values, we would obtain an overestimation of the CVA. Furthermore, this effect is likely to increase with decreasing credit quality of the counterparty. In practice, this means that the counterparty is paying a CVA which is based on a sub-optimal exercise strategy used by the bank, which is out of control for the counterparty. Even more problematic is that the overestimation of the CVA is higher for counterparties that already are under financial distress. 

One could argue that it is reasonable for the bank to charge the counterparty the higher CVA, since the bank will probably not follow the theoretically optimal risky exercise strategy. However, there is another level of complexity not yet discussed. When the mark-to-market (MtM) CVA moves in time against the bank, the bank could face losses, not because the counterparty actually defaults, but because disadvantageous changes in the MtM CVA. For instance, in Basel III \cite{BaselIII} the following is stated:

\vspace{0.2cm}
\noindent\say{\emph{Under Basel II, the risk of counterparty default and credit mitigation risk were addressed but mark-to-market losses due to credit valuation adjustments (CVA) were not. During the global financial crisis, however, roughly two-thirds of losses attributed to counterparty credit risk were due to CVA losses and only about one-third were due to actual defaults.}}
\vspace{0.2cm}

\noindent This is further discussed in \cite{Brigo}, in which the authors also recommend computations of different risk measures for the future distribution of CVA. Two examples of such measures are the Value at Risk of the CVA (VaR-CVA) and the Expected Shortfall of the CVA (ES-CVA). The advantage of the ES-CVA is that it is a coherent risk-measure, and we therefore focus on ES-CVA in this paper. 

\subsection{Structure of the paper}
In Section \ref{sec2} the mathematical problem formulation is given. We define the risk-free and risky portfolios, close-out agreements both with and without netting agreements and the associate CVA. Furthermore, the problems are formulated in terms of so-called decision functions, which control the exercise strategies. In Section \ref{sec3}, the algorithms are presented. In the first part, the algorithm for learning optimal exercise strategies is given and in the second part, an algorithm for learning pathwise entities such as the pathwise portfolio exposure is presented. Finally, in Section \ref{sec4} numerical experiments are presented. The experiments include a first part, in which risk-free values are computed and compared to a well-established regression based method. In the second part we compare CVA computed with the risk-free and the risky exercise strategy to verify that, indeed, the CVA is often overestimated with algorithms in use today. We present comparisons with and without netting, for different levels of Wrong Way Risk (WWR), and for different credit quality of the counterparty. As a final example, we analyse the effect of the different exercise strategies on ES-CVA.  In the Appendix, we provide some additional details on the algorithms and the specific choice of neural networks.

\section{Problem formulation}\label{sec2}
Let $(\Omega,\F,\Q)$ be a probability space completed with the $\Q-$null-sets of $\F$. For $T\in(0,\infty)$, and\footnote{We use $\N=\{1,2,3\ldots\}$ and $\N_0=\N\cup\{0\}$, and $\R_+=(0,\infty)$.} $d\in\N$, let $X\colon[0,T]\times\Omega\to\R^d$ and $r\colon[0,T]\times\Omega\to\R$ represent the (market) risk-factors of the portfolio and the short rate, respectively. Furthermore, we denote by $\tau^\text{\tiny{D}}$ the default event of the counterparty, which is a stopping time defined on $(\Omega,\F,\Q)$ and we let $\mathbbm{1}^\text{\tiny{D}}:[0,T]\times\Omega\to\{0,1\}$, be the \textit{jump-to-default} process given by \begin{align}
    \mathbbm{1}_t^\text{\tiny{D}}\coloneqq\I_{\{t<\tau^\text{\tiny{D}}\}}.
\end{align} 
The information structure is given by the sub$-\sigma-$algebras generated by $X$, $r$ and $\mathbbm{1}^\text{\tiny{D}}$, \textit{i.e.,} $\Ht^X=\sigma\left(X_s\colon s\in[0,t]\right)$, $\Ht^r=\sigma\left(r_s\colon s\in[0,t]\right)$ and $\mathcal{G}_t=\sigma\left(\mathbbm{1}_s^\text{\tiny{D}}\colon s\in[0,t]\right)$ and we define the enlarged filtrations $\Ht=\Ht^X\wedge\Ht^r$ and $\Ft=\Ht\wedge \mathcal{G}_t$. In this paper, we use either a constant short rate (risk-free rate), or we view the short rate as one of the risk factors. In the latter case, we model the short rate as one of the $d$ component processes of $X$, which implies that, $\Ht=\Ht^X$. The motivation for introducing a separate notation for the short rate is to simplify the notation when the short rate is used to discount cash-flows. For commonly used conditional expectations, we introduce the short-hand notations $\E_{t,x}\left[\,\cdot\,\right]\coloneqq\E
^\Q\left[\,\cdot\,|\,X_t=x\right]$, $\E_{t,x,\nu}\left[\,\cdot\,\right]\coloneqq\E
^\Q\left[\,\cdot\,|\,X_t=x,\mathbbm{1}_t^\text{\tiny{D}}=\nu\right]$ and $\E_{t}\left[\,\cdot\,\right]\coloneqq\E
^\Q\left[\,\cdot\,|\,\Ht\right]$. 

We use a numéraire, which, for $t\in[0,T]$, is defined by $B_t\coloneqq\ \text{exp}\left(\int_0^t r_s\d s\right)$, which 
should be interpreted as the value at time $t$ of a savings-account, which was worth 1 at time $0$. For $t,u\in[0,T]$ with $t\leq u$, we use $D_{t,u}\coloneqq\frac{B_t}{B_u}$ to discount a cash-flow obtained at time $u$ back to time $t$. The measure $\Q$ is the risk-free measure, under which all tradeable assets are martingales relative to the numéraire, \textit{e.g.,} if component $i\in\{1,2,\ldots,d\}$ of X is tradeable, then $\frac{(X_t)_i}{B_t}$ is a $\Q-$martingale.

If not specifically stated otherwise, equalities and inequalities of random variables should be interpreted in a $\Q-$almost sure sense.

\subsection{A portfolio of derivatives}\label{derivative_por}
We assume a portfolio of $J\in\N$ derivatives. For $t\in[0,T]$, and for derivative $j\in\{1,2,\ldots,J\}$, we denote the set of exercise dates greater than or equal to $t$ by $\mathbbm{T}_j(t)\subseteq[0,T]$, and set $\boldsymbol{\mathbbm{T}}(t)=\{\mathbbm{T}_1(t),\,\mathbbm{T}_2(t),\ldots,\mathbbm{T}_J(t)\}$. Note that for a European-type contract, the only exercise date is at the maturity, for a Bermudan-type contract there are multiple exercise dates, and for an American-type contract, there are infinitely many exercise dates. We emphasize that the exercise dates are simply subsets of the time interval $[0,T]$, and provide no information on which exercise policy to follow, except in some trivial cases \textit{e.g.,} when there is only one exercise date. 

Since we want to be able to treat derivatives with early-exercise features, we need to introduce a framework for stopping times. For $j\in\{1,2,\ldots,J\}$, an $X-$stopping time with respect to $\mathbbm{T}_j(0)$, is a random variable, $\tau_j$, defined on $(\Omega,\,\mathcal{F},\,\Q)$, taking on values in $\mathbbm{T}_j(0)$, such that for all $s\in\mathbbm{T}_j(t)$, it holds that the event $\{\tau_j = s\}\in\mathcal{H}_s$. Furthermore, we define an $X
^{t,x}-$stopping time as an $X-$stopping time, conditional on $X_t=x$, and $\tau\geq t$.

For each derivative, $j\in\{1,2,\ldots, J\}$ we use individual pay-off functions, $g_j\colon[0,T]\times\R^d\to\R$, which, for $t,s\in[0,T]$, with $s\geq t$, are assumed to satisfy \begin{equation}\label{assumtion2int}
    \E_0[|D_{t,s}g_j(s,X_s)|^2]<\infty.
\end{equation} 
Since we are treating portfolios where the individual derivatives may have different maturities, we set each pay-off function to zero for all times larger than its maturity, \textit{i.e.,} for $j\in\{1,2,\ldots,J\}$, $x\in\R^d$ and for $t>\max\{\mathbbm{T}_j(0)\}$, we set $g_j(t,x)\equiv0$, where $\max\{\mathbbm{T}_j(0)\}$ represents the largest element belonging to the set $\mathbbm{T}_j(0)$.   

\subsection{Risk-free and risky portfolio valuation without netting}\label{no_netting}
The value of a derivative (not) taking default risk of the counterparty into account is referred to as the \textit{risky} (\textit{risk-free}) value. We define the risk-free and the risky values of derivative $j\in\{1,2,\ldots,J\}$, at market state $(t\in[0,T],X_t=x\in\R^d)$, and default state $\mathbbm{1}_t^\text{\tiny{D}}=\nu\in\{0,1\}$, by \begin{align}\label{Vj}
 V_j(t,x)\coloneqq&\sup_{\tau\in{\mathcal{T}_j(t)}}\E_{t,x}\left[D_{t,\tau}g_j(\tau,X_\tau)\right],\quad \text{(risk-free value),}\\\begin{split}\label{Uj}
 U_j(t,x,\nu)\coloneqq&\nu\sup_{\tau\in{\mathcal{T}_j(t)}}\E_{t,x,1}\big[\mathbbm{1}_\tau^\text{\tiny{D}} D_{t,\tau}g_j(\tau,X_\tau) \\
 &+ (1-\mathbbm{1}_\tau^\text{\tiny{D}})D_{t,\tau^\text{\tiny{D}}}\left(RV_j(\tau^\text{\tiny{D}},X_{\tau^\text{\tiny{D}}})^++V_j(\tau^\text{\tiny{D}},X_{\tau^\text{\tiny{D}}})^-\right)\big],    \quad \text{(risky value),}\end{split}
\end{align}
where $\mathcal{T}_j(t)$ is the set of all $X-$stopping times taking on values in $\mathbbm{T}_j(t)$ and for $x\in\R$, $(x)^+=\max\{0,x\}$ and $(x)^-=\min\{0,x\}$. In the above, we assume a close-out agreement which uses the risk-free derivative values as reference valuation. At default of the counterparty, the bank receives only a fraction, $R\in[0,1)$, referred to as the \textit{recovery-rate}, of the positive part of each derivative. On the other hand, each derivative with a negative risk-free value at default needs to be added entirely to the portfolio. 

Note that for the risky value we need additional information of prior defaults of the counterparty, which is captured in the realization, $\nu\in\{0,1\}$, of the jump-to-default process, \textit{i.e.,} $\nu=1$ if no default has occurred prior to, or at $t$, and $\nu=0$ otherwise. The notation above trivially holds for European-type derivatives since the only exercise date is at maturity of the contract. Furthermore, a barrier-type feature could be added by also including a spatial dimension to $\mathbbm{T}_j(0)$. The value of a portfolio, consisting of $J$ derivatives, at market state $(t,X_t=x)$ and default state $\mathbbm{1}_t^\text{\tiny{D}}=\nu$, without netting, is given by 
\begin{equation*}
    \Pi^\textit{\tiny{V}}(t,x)\coloneqq\sum_{j=1}^{J}V_j(t,x), \quad \Pi^{\textit{\tiny{U}}}(t,x,\nu)\coloneqq\sum_{j=1}^{J}U_j(t,x,\nu)=\nu\sum_{j=1}^{J}U_j(t,x,1).
\end{equation*}
Using \eqref{Vj} and \eqref{Uj}, the above can be written as\begin{align}\label{Vtauportfolio}
    \Pi^\textit{\tiny{V}}(t,x)=&\sum_{j=1}^J\sup_{\tau_j\in\T_j(t)}\E_{t,x}\big[D_{t,\tau_j}g_j(\tau_j,X_{\tau_j})\big]\\\begin{split} \Pi^\textit{\tiny{U}}(t,x,\nu)=&
    \nu\sum_{j=1}^J\sup_{\tau_j\in\T_j(t)}\E_{t,x,1}\Big[\mathbbm{1}_{\tau_j}^\text{\tiny{D}}D_{t,\tau_j}g_j(\tau_j,X_{\tau_j}) \\
    &+ (1-\mathbbm{1}_{\tau_j}^\text{\tiny{D}})D_{t,\tau^\text{\tiny{D}}}\big(RV_j(\tau^\text{\tiny{D}},X_{\tau^\text{\tiny{D}}})^++V_j(\tau^\text{\tiny{D}},X_{\tau^\text{\tiny{D}}})^-\big)\Big].\label{Utauportfolio}\end{split}
\end{align}
Since the aim is to approximate the optimal exercise policy with neural networks, we wish to re-formulate the problem into an optimization problem, in which the target function can be represented by a neural network. Following \cite{DOS} and \cite{DOSexposure} we use so-called decision functions, to determine for each derivative and given a market state, whether or not to exercise the derivative. For $j\in\{1,2,\ldots,J\}$, decision function $j$ denoted by $f_j$, is of the form $f_j\colon[0,T]\times\R^d\to\{0,1\}$. In order to guarantee that an exercise decision can only occur at an exercise date, we require for $s\notin\mathbbm{T}_j(0)$, that $f_j(s,\cdot)\equiv 0$.

We now restrict our attention to the case when there, for each derivative, is a finite number of exercise dates, $\textit{i.e.,}$ for $j\in\{1,2,\ldots,J\}$, it holds that $|\mathbbm{T}_j(0)|\in\N$. From a theoretical perspective, this excludes American-type derivatives, but from a practical perspective, an infinite number of exercise dates is often approximated by a large, but finite, number of exercise dates. This implies that we can still consider American-type derivatives by increasing the number of exercise dates until the derivative value converges (until the value does not increase with additional exercise dates). We denote by $\mathbbm{T}^\Pi(t)$ the set of dates which represent an exercise date for at least one of the $J$ derivatives. Mathematically, we define the exercise dates of the portfolio as
\begin{equation}\label{unionExDates}
    \mathbbm{T}^\Pi(t)\coloneqq\bigcup_{j=1}^J\mathbbm{T}_j(t),
\end{equation}
and the number of unique exercise dates in the portfolio is given by $N=|\mathbbm{T}^\Pi(0)|$. We assume that the initial time $T_0=0$ is not an exercise date for any of the derivatives.

To simplify, we use the following notation for the $N$ exercise dates, the risk-factors evaluated at the $N$ exercise dates, and the discounting between exercise dates 
\begin{align}\label{TPortfolio}
    \mathbbm{T}^\Pi(0)&=\{T_1,T_2,\ldots,T_N=T\},\\
    X_k&\coloneqq X_{T_k},\text{ and }D_{k,\ell}\coloneqq D_{T_k,T_{\ell}}\quad\text{for } k,\ell=1,2,\ldots,N.\label{RF_atexd}
\end{align} 
Furthermore, for $t\in[0,T]$, the risk-factor process on $[t,T]$, conditional on $X_t=x$, is denoted by $X^{t,x}=(X_s)_{s\in[t,T]}$, where we also note that $X^{0,x_0}=X$. 
The above notation allows us to express an $X^{t,x}-$stopping time in terms of decision functions\footnote{The empty product is defined as 1.}
\begin{equation}\label{taufstar}
\tau_j^n[f_j](X^{t,x})\coloneqq\sum_{k=n}^{N}T_kf_j(T_k,X_k)\prod_{m=n}^{k-1}(1-f_j(T_m,X_m)).    
\end{equation}
The notation above is used to emphasize that, the decision function $f_j$, controls the exercise strategy, given the stochastic process $X^{t,x}$. Moreover, $X^{t,x}$ is not just a random value at a specific time, but the entire process, starting at $X_t=x$ and until stopping occurs. In later sections the valuation of a derivative or a portfolio is formulated as an optimization problem, which is optimized by varying $f_j$. Although the notation is practical when optimization is discussed, it is cumbersome to use when we define the value of a derivative. We therefore use the following short-hand notation
\begin{equation}\label{taustarf_short}
    \bar{\tau}_{n,j}\coloneqq\tau_j^n[f_j](X^{t,x}),
\end{equation}
and keep in mind, that the strategy is controlled by a decision function, $f_j$, and for $u\geq t$, the event $\I_{\{\bar{\tau}_{n,j}\leq u\}}$ is $\sigma(X^{t,x})-$measurable.
We can now define the value of the risk-free and risky derivatives, given an exercise strategy expressed in terms of decision functions. For derivative $j\in\{1,2,\ldots,J\}$, market state $(t,x)\in(T_{n-1},T_n]\times\R^d$, we define the parametrized valuation functions
\begin{align}\label{curlyVj}
    \mathcal{V}_j(t,x\,|\,f_j)\coloneqq&\E_{t,x}\left[D_{t,\bar{\tau}_{n,j}}\,g_j\left(\bar{\tau}_{n,j},X_{\bar{\tau}_{n,j}}\right)\right],\\\begin{split}\label{curlyUj}
    \mathcal{U}_j(t,x\,|\,f_j)\coloneqq&\E_{t,x,1}\big[\mathbbm{1}_{\bar{\tau}_{n,j}}^\text{\tiny{D}}D_{t,\bar{\tau}_{n,j}}g_j\big(\bar{\tau}_{n,j},X_{\bar{\tau}_{n,j}}\big)+ \big(1-\mathbbm{1}_{\bar{\tau}_{n,j}}^\text{\tiny{D}}\big)D_{t,\tau^\text{\tiny{D}}}\big(RV_j(\tau^\text{\tiny{D}},X_{\tau^\text{\tiny{D}}})^++V_j(\tau^\text{\tiny{D}},X_{\tau^\text{\tiny{D}}})^-\big)\big].\end{split}
\end{align}
Similarly, we define the portfolio values with respect to the exercise strategy given by $\boldsymbol{f}$, as the parametrized functions
\begin{equation}\label{upsilon}
    \Upsilon^\textit{\tiny{V}}\left(t,x\,|\,\boldsymbol{f}\right)\coloneqq\sum_{j=1}^J\mathcal{V}_j(t,x\,|\,f_j),\quad
     \Upsilon^\textit{\tiny{U}}\left(t,x,\nu\,|\,\boldsymbol{f}\right)\coloneqq\nu\sum_{j=1}^J\mathcal{U}_j(t,x\,|\,f_j),
\end{equation}
where the value of the risky portfolio also depends on the default state of the counterparty, $\mathbbm{1}_t^\text{\tiny{D}}=\nu\in\{0,1\}$. We now want to find decision functions such that, when inserted in \eqref{curlyVj} and \eqref{curlyUj}, we obtain \eqref{Vj} and \eqref{Uj}. With this in mind, we define for $j\in\{1,2,\ldots,J\}$, at $t\in[0,T]$, the (optimal) exercise regions, $\mathcal{E}^\textit{\tiny{Z}}_j(t)$, in which it is optimal to exercise, and the (optimal) continuation regions, $\mathcal{C}^\textit{\tiny{Z}}_j(t)$, in which it is optimal to hold on, by
\begin{align*}
    \mathcal{E}^\textit{\tiny{Z}}_j(t) \coloneqq& \left\{x\in\mathbbm{R}^d\,|\,Z_j(t,x)=g_j(t,x)\,\text{ and }\,t\in \mathbbm{T}_j(0)\right\},\\
    \mathcal{C}^\textit{\tiny{Z}}_j(t) \coloneqq& \left\{x\in\mathbbm{R}^d\,|\,Z_{j}(t,x)>g_{j}(t,x)\,\text{ or }\,t\notin \mathbbm{T}_j(0)\right\}, \ \text{for } Z\in\{V,U\}.
\end{align*}
The above states that the derivative should be exercised if its value equals the immediate exercise value, and we are at an exercise date and the derivative should not be exercised if its value is greater than the immediate exercise value or if we are not at an exercise date. Note that $\mathcal{E}^\textit{\tiny{Z}}_j(t)\cup\mathcal{C}^\textit{\tiny{Z}}_j(t)=\R^d$ and $\mathcal{E}^\textit{\tiny{Z}}_j(t)\cap\mathcal{C}^\textit{\tiny{Z}}_j(t)=\emptyset$. For $t\in[0,T]$, a decision function, $j\in\{1,2,\ldots,J\}$, can then be defined as
\begin{equation}\label{fZstar}
    f_j^\textit{\tiny{Z}}(t,x)\coloneqq\I_{\{x\in\mathcal{E}^\textit{\tiny{Z}}_j(t)\}},\ \text{for } Z\in\{V,U\}.
\end{equation}
Furthermore, we denote by $\boldsymbol{f}^\textit{\tiny{Z}}$, the vector consisting of the individual decision functions \begin{equation}\label{boldfZstar}
    \boldsymbol{f}^\textit{\tiny{Z}}(t,x)\coloneqq(f_1^\textit{\tiny{Z}}(t,x),f_2^\textit{\tiny{Z}}(t,x),\ldots,f_J^\textit{\tiny{Z}}(t,x))^T, \  \text{for } Z\in\{V,U\}.
\end{equation}
For market states $(t,x)\in[0,T]\times\R^d$ and for a derivative $j\in\{1,2,\ldots,J\}$, it holds that
\begin{equation*}
    \mathcal{V}_j\big(t,x\,|\,f^D_j,\,f^{\text{\tiny{D}}}_j\big)=V_j(t,x),\quad \mathcal{U}_j\big(t,x\,|\,f^\textit{\tiny{U}}_j\big)=U_j(t,x,1).
\end{equation*}
The validity of the above is a direct consequence of Proposition 4 in \cite{DOS}. In turn, this implies that by inserting the optimal decision functions in the functionals in equations \eqref{upsilon}, we obtain the risky and risk-free portfolio values, \textit{i.e.},
\begin{equation*}
    \Upsilon^{V}\big(t,x\,|\,\boldsymbol{f}^\textit{\tiny{V}}\big)=\Pi^\textit{\tiny{V}}(t,x), \quad \Upsilon^{
    U}\big(t,x,\nu\,|\,\boldsymbol{f}^\textit{\tiny{U}}\big)=\Pi^\textit{\tiny{U}}(t,x,\nu).
\end{equation*}

In subsequent sections the optimal decision function $\boldsymbol{f}^\textit{\tiny{Z}}$, for $Z\in\{V,U\}$, is approximated with a series of neural networks. The reason for using the rather complicated notation, \eqref{taufstar}, is that this structure allows us to view the valuation of the derivatives as an optimization problem over the set of decision functions, which we approximate on some finite-dimensional function space. One example of such function space is the functions generated by a series of neural networks with a fixed number of parameters. When we use a specific strategy, \textit{e.g.,} $\boldsymbol{f}^\textit{\tiny{V}}$ or $\boldsymbol{f}^\textit{\tiny{U}}$, this is specified by adding a superscript referring to the particular strategy. For $Z\in\{V,U\}$, we define the short-hand notation \begin{equation}\label{shorttaustar}
    \bar{\tau}_{n,j}^\textit{\tiny{Z}}\coloneqq\tau_j^n[f^\textit{\tiny{Z}}_j](X^{t,x}),
\end{equation} 
where it is assumed that $t\in(T_{n-1},T_n]$.

\subsection{Risky portfolio valuation with netting}
When considering the risky portfolio value with netting, the problem becomes nonlinear in the sense that the risky portfolio value is no longer the sum of the individual risky derivative values. In fact, there no longer exists ''a risky value for a single derivative'', since the valuation needs to be carried out on a portfolio level. Before we define the risky value of a netted portfolio, we need to define the process\footnote{$\{0,\,1\}^J$ is the Cartesian product of $\{0,\,1\}\times\{0,\,1\}\times\cdots\times\{0,\,1\}$, $J$ times.} $A\colon[0,T]\times\Omega\to\{0,1\}^J$, which for $t\in[0,T]$ and $j\in\{1,2,\ldots,J\}$ satisfies 
\begin{equation*}
    (A_t)_j=\begin{cases}0,\quad\text{ if derivative $j$ has been exercised prior to $t$,}\\
    1,\quad\text{ else.}\end{cases}
\end{equation*}
Similar to \eqref{RF_atexd}, we use the short-hand notation for $A$ at initial date, $T_0$, the exercise dates, $T_1,\ldots,T_N$,
\begin{equation}
    A_k \coloneqq A_{T_k},\quad \text{for }k=0,1,\ldots,N.
\end{equation}
The process $A$ is $\Ht-$measurable but it is not enough to know $X_t=x$ in order to determine $A_t$. The reason for defining $A$ is that the exercise decisions for the netted risky portfolio are defined by a $J-$dimensional $(X^{t,x},A^{t,\alpha})-$stopping times vector, where $A^{t,\alpha}=(A_s)_{s\in[t,T]}$ conditional on $A_t=\alpha$. This means that, at each exercise date, in addition to the current market state, we need to know which derivatives in the portfolio have been exercised prior to the current time, in order to make optimal exercise decisions. We denote, by $\boldsymbol{\mathcal{T'}}(t)$, the space of $(X^{t,x},A^{t,\alpha})-$stopping times vectors, taking on values in $\{\mathbbm{T}_1(t),\mathbbm{T}_2(t),\ldots,\mathbbm{T}_J(t)\}$. Furthermore, for $t\in(T_{n-1},T_n]$, we denote by $\tau_j^n$ element $j$ of a stopping times vector
$\boldsymbol{\tau}^n\in\boldsymbol{\mathcal{T'}}(t)$. The netted portfolio value with a risky counterparty, given market state $X_t=x$, default state of the counterparty $\mathbbm{1}_t^\text{\tiny{D}}=\nu$ and portfolio state (exercise state of the derivatives in the portfolio) $A_t=\alpha$, is given by
\begin{align}\begin{split}\label{U2tauportfolio}
    \Pi^\textit{\tiny{A}}(t,x,\nu,\alpha)\coloneqq&\nu\sup_{\boldsymbol{\tau}\in\boldsymbol{\T'}(t)}\E_{t,x,1,a}\Bigg[\sum_{j=1}^J\alpha_j\mathbbm{1}_{\tau_j}^\text{\tiny{D}}D_{t,\tau_j}g_k(\tau_j,X_{\tau_j})\\
    & +D_{t,\tau^\text{\tiny{D}}}\Bigg(R\Bigg(\sum_{k=1}^J\alpha_k(1-\mathbbm{1}_{\tau_k}^\text{\tiny{D}})V_k(\tau^\text{\tiny{D}},X_{\tau^\text{\tiny{D}}})\Bigg)^+
    +\Bigg(\sum_{\ell=1}^J\alpha_\ell(1-\mathbbm{1}_{\tau_\ell}^\text{\tiny{D}})V_j(\tau^\text{\tiny{D}},X_{\tau^\text{\tiny{D}}})\Bigg)^-\Bigg)\Bigg],\end{split}
\end{align}
where $\E_{t,x,\nu,\alpha}[\,\cdot\,]=\E^\Q[\,\cdot\,|\,X_t=x,\mathbbm{1}_t^\text{\tiny{D}}=\nu,A_t=\alpha]$.
To emphasize the importance, we put the following observations about \eqref{U2tauportfolio} into two remarks. 
\begin{remark}
The optimal stopping strategies of the individual derivatives in the portfolio are no longer independent of each other, as in \eqref{Vtauportfolio} and \eqref{Utauportfolio}. Furthermore, the optimal strategy depends on earlier exercise decisions, meaning that in order to make the exercise decisions Markovian, we need to include information about earlier decisions. The reason for this is the non-linearity in the two sums inside the expectation in \eqref{U2tauportfolio}. Therefore, the optimal stopping strategies need to be computed for the entire portfolio simultaneously. To the best of our knowledge, this has not been done in an ordinary least squares setting before. However, it is discussed in a PDE framwork in the case of a portfolio of American swaptions in \cite{CVA_BER3}.
\end{remark}

\begin{remark}
The value of the risky portfolio with netting, depends on the $J$ risk-free derivative values and we are therefore required to approximate the risk-free derivative values. The reason for this is that, at default, the risk-free value of the portfolio is used as reference value in the close-out agreement (see Equation \eqref{U2tauportfolio}). If we restrict our portfolio to derivatives with positive pay-off functions, then $V_j$ can be replaced by $g_j$ (by the definition of $V_j$ and the law of iterated expectations). Furthermore, in the restricted portfolio, the values with and without netting coincide.
\end{remark}
An $(X^{t,x},A^{t,\alpha})-$stopping times vector can be defined by 
\begin{equation}\label{taufstar_net}
\boldsymbol{\tau}^n[\boldsymbol{f}](X^{t,x},A^{t,x})\coloneqq\sum_{k=n}^NT_k\boldsymbol{f}(T_k,X_k,A_k)\odot \prod_{M=n}^{k-1}(\boldsymbol{1}_J-\boldsymbol{f}(T_m,X_m,A_m)),
\end{equation}
where $\odot$ is element-wise multiplication and $\boldsymbol{1}_J$ is the $J-$dimensional vector with only ones, $(1,1,\ldots,1)^T$. We denote element $j$ of the stopping times vector by $\tau^n_j[\boldsymbol{f}](X^{t,x},A^{t,\alpha})=\left(\boldsymbol{\tau}^n[\boldsymbol{f}](X^{t,x},A^{t,x})\right)_j$. We emphasize that each element of the stopping time vector depends on $\boldsymbol{f}$ and not only an element $j$ which is the case without netting. Similar to \eqref{taustarf_short}, we introduce a short-hand notation, which simplifies the valuation function\begin{equation}
    \hat{\boldsymbol{\tau}}_n\coloneqq\boldsymbol{\tau}^n[\boldsymbol{f}](X^{t,x},A^{t,x}),\quad \hat{\tau}_{n,j}\coloneqq(\hat{\boldsymbol{\tau}}_n)_j.
\end{equation}
We here use ''$\hat{\tau}$ '', instead of ''$\bar{\tau}$'' as in \eqref{taustarf_short}, to emphasize that the stopping time also takes $A^{t,\alpha}$ as an argument. The netted risky portfolio value, given the exercise strategy obtained by decision function $\boldsymbol{f}$, is then given by
\begin{align}\begin{split}\label{upsilon2}
           \Upsilon^\textit{\tiny{A}}\big(t,x,&\nu,\alpha\,|\,\boldsymbol{f}\big)\coloneqq\nu\E_{t,x,1,\alpha}\Bigg[\sum_{j=1}^J\alpha_j\mathbbm{1}_{\hat{\tau}_{n,j}}^\text{\tiny{D}}D_{t,\hat{\tau}_{n,j}}g_j\big(\hat{\tau}_{n,j},X_{\hat{\tau}_{n,j}}\big)\\
           &+D_{t,\tau^\text{\tiny{D}}}\Bigg(R\Bigg(\sum_{j=1}^J\alpha_j(1-\mathbbm{1}_{\hat{\tau}_{n,j}}^\text{\tiny{D}})V_j(\tau^\text{\tiny{D}},X_{\tau^\text{\tiny{D}}})\Bigg)^++\Bigg(\sum_{j=1}^J\alpha_j(1-\mathbbm{1}_{\hat{\tau}_{n,j}}^\text{\tiny{D}})V_j(\tau^\text{\tiny{D}},X_{\tau^\text{\tiny{D}}})\Bigg)^-\Bigg)\Bigg],\end{split}
\end{align}
where we, again, remind ourselves that the exercise strategy is controlled by $\boldsymbol{f}$, and for $u\geq t$, the event $\I_{\{\hat{\tau}_{t,j}\leq u\}}$ is $\sigma(X^{t,x},A^{t,\alpha})-$measurable.

For a netted portfolio, the optimal exercise regions, described in Section \ref{no_netting}, are less trivial. Firstly, they become dependent on the state of earlier exercise decisions, $A_t=\alpha_t\in\{0,1\}^J$. Secondly, the exercise region for derivative $j\in\{1,2,\ldots,J\}$ is expressed under the condition that an optimal exercise strategy for the other $J-1$ derivatives is applied. Therefore, we only describe the optimal decision function as belonging to the supremum over the space, $\mathcal{D}$, of all measurable functions, $f\colon[0,T]\times\R^d\times\{0,1\}^J\to\{0,1\}^J$,
\begin{equation}
    \boldsymbol{f}^\textit{\tiny{A}}\in\argmax_{\boldsymbol{f}\in\mathcal{D}}\Upsilon^\textit{\tiny{A}}\big(0,x_0,\nu_0,\alpha_0\,|\,\boldsymbol{f}\big),
\end{equation}
where $\nu_0=1$ (no default prior to or at $t=0$) and $a_0=(1,1,\ldots,1)^T$ (no derivatives have been exercised prior to $t=0$). We then assume that, given the state $(t,X_t=x,\mathbbm{1}_t^\text{\tiny{D}}=\nu,A_t=\alpha)$, the following holds
\begin{equation}\label{upsilonPi}
    \Upsilon^\textit{\tiny{A}}(t,x,\nu,\alpha\,|\,\boldsymbol{f}^\textit{\tiny{A}})=\Pi^\textit{\tiny{A}}(t,x,\nu,\alpha).
\end{equation}
Similar to \eqref{shorttaustar}, when we want to emphasize the particular choice of decision function, $\boldsymbol{f}^\textit{\tiny{A}}$, we use the short-hand notation \begin{equation}\label{shorttaustar_A}
    \hat{\boldsymbol{\tau}}_n^\textit{\tiny{A}}=\boldsymbol{\tau}^n[\boldsymbol{f}^\textit{\tiny{A}}](X^{t,x},A^{t,\alpha}) \quad\text{and}\quad  \hat{\tau}_{n,j}^\textit{\tiny{A}}=\left(\boldsymbol{\tau}^n[\boldsymbol{f}^\textit{\tiny{A}}](X^{t,x},A^{t,\alpha})\right)_j,
\end{equation} 
where it is assumed that $t\in(T_{n-1},T_n]$.

\subsection{Credit valuation adjustment of a derivative portfolio}
The formal definition of CVA is the difference between the risk-free and the risky portfolio value. Given models of the underlying market and default events of our counterparty, the above definition of CVA is straight-forward for a portfolio consisting of derivatives without optionality \textit{e.g.,} European options, barrier options etc. When it comes to portfolios consisting of derivatives with true optionality, \textit{e.g.,} the Bermudan options, American options etc. the standard procedure is not clear. In this section, we define the CVA for portfolios of derivatives with true optionality as well as some approximations, which simplify the computations. In the definitions of CVA, we use the portfolio valuations in terms of optimally chosen decision functions given in equations \eqref{upsilon} and \eqref{upsilonPi}. The CVA at $(t=0,X_0=x_0)$, with and without netting, respectively, are given by
\begin{align}\label{CVA_no_net}
    \text{CVA}\coloneqq\Upsilon^\textit{\tiny{V}}\big(0,x_0\,|\,\boldsymbol{f}^\textit{\tiny{V}}\big)-\Upsilon^\textit{\tiny{U}}\big(0,x_0,1\,|\,\boldsymbol{f}^\textit{\tiny{U}}\big),\quad(\text{without netting}),\\
    \text{CVA}^\text{Net}\coloneqq\Upsilon^\textit{\tiny{V}}\big(0,x_0\,|\,\boldsymbol{f}^\textit{\tiny{V}}\big)-\Upsilon^\textit{\tiny{A}}\big(0,x_0,1,\boldsymbol{1}_J\,|\,\boldsymbol{f}^\textit{\tiny{A}}\big),\quad(\text{with netting}).\label{CVA_net}
\end{align}
A commonly used approximation is to apply the same exercise strategy to the risk-free and risky portfolios. One such approximation is defined as
\begin{align}\label{CVA_no_net_approx}
    \overline{\text{CVA}}\coloneqq\Upsilon^\textit{\tiny{V}}\big(0,x_0\,|\,\boldsymbol{f}^\textit{\tiny{V}}\big)-\Upsilon^\textit{\tiny{U}}\big(0,x_0,1\,|\,\boldsymbol{f}^\textit{\tiny{V}}\big),\quad(\text{Risk-free strategy, without netting}),\\
    \overline{\text{CVA}}^\text{Net}\coloneqq\Upsilon^\textit{\tiny{V}}\big(0,x_0\,|\,\boldsymbol{f}^\textit{\tiny{V}}\big)-\Upsilon^\textit{\tiny{A}}\big(0,x_0,1,\boldsymbol{1}_J\,|\,\boldsymbol{f}^\textit{\tiny{V}}\big),\quad(\text{Risk-free strategy, with netting}).\label{CVA_net_approx}
\end{align}
The only difference between \eqref{CVA_no_net}-\eqref{CVA_net} and \eqref{CVA_no_net_approx}-\eqref{CVA_net_approx} is that in the latter the risk-free strategy is used also for the risky portfolios. One could also think of other definitions, \textit{e.g.,} using the risky strategies for both portfolios. This particular choice is motivated by the fact that $\boldsymbol{f}^\textit{\tiny{U}}$ and $\boldsymbol{f}^\textit{\tiny{A}}$ are, in general, dependent on $\boldsymbol{f}^\textit{\tiny{V}}$ through the close-out agreements in \eqref{curlyUj} and \eqref{upsilon2}. Moreover, $\boldsymbol{f}^\textit{\tiny{V}}$ is a sub-optimal strategy for both risky portfolios leading to $\text{CVA}\leq\overline{\text{CVA}}$, and $\text{CVA}^{\text{Net}}\leq\overline{\text{CVA}}^{\text{Net}}$, which is beneficial for the bank (but certainly not for the counterparty). If we instead use only the risky descision functions, \textit{i.e.,} replacing $\boldsymbol{f}^\textit{\tiny{V}}$ with $\boldsymbol{f}^\textit{\tiny{U}}$ in \eqref{CVA_no_net_approx} and $\boldsymbol{f}^\textit{\tiny{A}}$ in \eqref{CVA_net_approx}, we would obtain an underestimation of the CVAs, which would be unacceptable for the bank.

As mentioned in the Introduction, the bank is exposed to the risk of CVA losses, as a consequence of the MtM value of the CVA moving against the bank. We therefore want to follow the evolution of the CVA over time, to gain insights in its distribution. Of particular interest is the tail distribution of the CVA, for times between initial time and the maturity of the portfolio. To explore this, we define the dynamic versions of \eqref{CVA_no_net}-\eqref{CVA_net_approx}, which are stochastic processes depending on the market and portfolio state processes $X$ and $A$. For $t\in[0,T]$, the dynamic versions of the CVAs (and their approximations) are given by the following random variables

\begin{align*}
\text{CVA}(t,X_t,A_t)\coloneqq&\sum_{j=1}^J\big(\mathcal{V}_j\big(t,X_t\,|\,f_j^\textit{\tiny{V}}\big)-\mathcal{U}_j\big(t,X_t\,|\,f_j^\textit{\tiny{U}}\big)\big)(A_t)_j,\\   
\text{CVA}^{\text{net}}\big(t,X_t,A_t)\coloneqq&\sum_{j=1}^J\mathcal{V}_j(t,X_t\,|\,f_j^\textit{\tiny{V}}\big)(A_t)_j-\Upsilon^\textit{\tiny{A}}\big(t,X_t,1,A_t\,|\,\boldsymbol{f}^\textit{\tiny{A}}\big),\\
\overline{\text{CVA}}(t,X_t,A_t)\coloneqq&\sum_{j=1}^J\big(\mathcal{V}_j\big(t,X_t\,|\,f_j^\textit{\tiny{V}}\big)-\mathcal{U}_j\big(t,X_t\,|\,f_j^\textit{\tiny{V}}\big)\big)(A_t)_j,\\
\overline{\text{CVA}}^\text{net}(t,X_t,A_t)\coloneqq&\sum_{j=1}^J\mathcal{V}_j(t,X_t\,|\,f_j^\textit{\tiny{V}}\big)(A_t)_j-\Upsilon^\textit{\tiny{A}}\big(t,X_t,1,A_t\,|\,\boldsymbol{f}^\textit{\tiny{V}}\big).
\end{align*}
In the above, the CVA is conditional on that the counterparty has not defaulted prior to, or at, $t$ (it does not make sense to calculate the CVA if the counterparty has already defaulted). From the above we can define the Expected value of the CVA (E-CVA), and for $\alpha\in(0,1)$, the $\alpha-$level of Value at Risk of the CVA (VaR-CVA) and Expected Shortfall of the CVA (ES-CVA), 
\begin{align}
    \label{ECVA}
        \text{E-CVA}(t) &\coloneqq \E\left[\text{CVA}(t,X_t,A_t)\,|\,\mathbbm{1}_t=1\right],\\
        \label{VaRCVA}
    \text{VaR-CVA}_\alpha(t) &\coloneqq \inf\Big\{ P\in\R \,\big|\,\Q\big(\text{CVA}(t,X_t,A_t)\leq P\big)\geq\alpha\Big\},\\ \label{ESCVA}
    \text{ES-CVA}_\alpha(t) &\coloneqq\E\big[\text{CVA}(t,X_t,A_t)\,\big|\,\mathbbm{1}_t=1\,,\,\text{CVA}(t,X_t,A_t)\geq\text{VaR-CVA}_\alpha(t) \big].
\end{align}
In a similar way $\text{E-}\overline{\text{CVA}}(t)$, $\text{ES-}\overline{\text{CVA}}_\alpha(t)$, $\text{E-CVA}^\text{net}(t)$, $\text{ES-CVA}^\text{net}_\alpha(t)$, $\text{E-}\overline{\text{CVA}}^\text{net}(t)$ and $\text{ES-}\overline{\text{CVA}}_\alpha^\text{net}(t)$ are defined. The expression for the ES-CVA looks complicated but is basically just the expected value of the $\alpha-$tail of the CVA distribution. We focus on ES-CVA instead of VaR-CVA because it is a coherent risk measure and VaR-CVA is not.
\begin{remark}
Since  $\mathrm{ES}$-$\mathrm{CVA}$ is a non-traded risk measure, it should ideally be computed under the real world measure $\mathbbm{P}$, see \textit{e.g.,} \cite{Brigo} for a detailed discussion. To be precise, $(X_t,A_t)$ should be generated under the $\mathbbm{P}-$measure and, the $\mathrm{CVA}$, which is a tradeable asset, should be computed under the $\Q-$measure. It is straight-forward to adjust the algorithms in this paper be able to compute $\mathrm{ES}$-$\mathrm{CVA}$ under the $\P-$measure, see \cite{DOSexposure} for details in the special case $J=1$.
\end{remark}

\subsection{Exposure profiles}
In this subsection we discuss the concept of exposure profiles for a portfolio of derivatives. The financial exposure (of the bank) is defined as the maximum amount the bank stands to loose if the counterparty defaults. The exposure profile is loosely defined as the distribution of the exposure over time. The exposures, with and without netting, are defined as
\begin{equation*}
    \text{E}^{\text{Net}}_t\coloneqq\max\bigg\{\sum_{j=1}^JV_j(t,X_t)(A_t)_j,\,0\bigg\},\quad \text{E}_t\coloneqq\sum_{j=1}^J\max\big\{V_j(t,X_t)(A_t)_j,\,0\big\},
\end{equation*}
where we recall that $(A_t)_j=\I_{\{\tau_j>t\}}$ with $\tau_j$ being the exercise date for derivative $j$. Furthermore, for a portfolio without netting, the expected exposure (EE), and for $\alpha\in(0,1)$, the potential future exposure (PFE) are defined as
\begin{align}
    \label{EE}
        \text{EE}(t) &\coloneqq \E_0\big[D_{0,t}\,\text{E}_t\big],\\
        \label{PFE}
    \text{PFE}_\alpha(t) &\coloneqq \inf\big\{P\in\R \,\big|\,\Q\big(D_{0,t}\,\text{E}_t\leq P\big)\geq\alpha\big\}.
\end{align}
Both the expectation and the probability in \eqref{EE} and \eqref{PFE} should be interpreted as conditional on $X_0=x_0\in\R^d$.
The EE and PFE in the presence of netting, denoted by $\text{EE}^\text{Net}(\cdot)$ and $\text{PFE}_\alpha^\text{Net}(\cdot)$, and are obtained by instead using the netted exposure in \eqref{EE} and \eqref{PFE}. 

If we assume a constant recovery rate $R\in[0,1)$, and that $X$ and $\mathbbm{1}^\text{\tiny{D}}$ are independent, \textit{i.e.,} the default event of the counterparty is independent of the risk factors, then \eqref{CVA_no_net_approx} and \eqref{CVA_net_approx} can be written as
\begin{equation*}
    \overline{\text{CVA}}=(1-R)\int_0^{T}\text{EE}(t)\Q\left(\tau^\text{\tiny{D}}\in [t+\d t)\right), \quad     \overline{\text{CVA}}^{\text{Net}}=(1-R)\int_0^{T}\text{EE}^\text{Net}(t)\Q\left(\tau^\text{\tiny{D}}\in [t+\d t)\right),    
\end{equation*}
which can be approximated as 
\begin{equation*}
    \overline{\text{CVA}}\approx(1-R)\sum_{m=1}^{M}\text{EE}(t_m)\Q\left(\tau^\text{\tiny{D}}\in (t_{m-1}, t_{m}]\right), \      \overline{\text{CVA}}^{\text{Net}}\approx(1-R)\sum_{m=1}^{M}\text{EE}^\text{Net}(t_m)\Q\left(\tau^\text{\tiny{D}}\in (t_{m-1}, t_{m}]\right),    
\end{equation*}
for some partition of $[0,T]$, with $t_0=0$ and $t_M=T$. The above formulations require access to the density of default events, but may be more accurate, especially for large $M$ and a small probability of default (with a simulation based approach, problems with a low probability of default can often be tackled with variance reduction techniques).

\section{Algorithms}\label{sec3}
In the first part of this section, we present a neural network-based method to approximate the decision functions introduced in the previous section. The method generalizes the Deep Optimal Stopping proposed in \cite{DOS} and extended in \cite{DOSexposure}, which approximates stopping decisions for a single derivative, to be applicable also for portfolios of derivatives with early-exercise features. Furthermore, for the risky portfolios, the algorithm is extended to be able to deal with default risk of the counterparty. The algorithm is based on a series of neural networks, which are optimized backwards in time with the objective to maximize the expected discounted cash-flows. 

In the second part of this section, the exercise policy obtained from the approximate decision functions is applied pathwise on realizations of the risk factors of each derivative in the portfolio to generate pathwise cash-flows. These cash-flows are used in a neural network based regression algorithm to approximate pathwise derivative values. These pathwise derivative values can then be used to compute important risk management measures.

\subsection{Phase I: Learning exercise strategy}\label{DOSalgo}
As indicated above, the core of the algorithm is to approximate decision functions, in order to obtain good approximations of the value of a portfolio of derivatives. We approximate the decision functions $\boldsymbol{f}^\textit{\tiny{V}}$, $\boldsymbol{f}^\textit{\tiny{U}}$ and $\boldsymbol{f}^\textit{\tiny{A}}$, with fully connected neural networks. To be more precise, let $N=|\mathbbm{T}^{\Pi}(0)|$, for $n\in\{1,2,\ldots,N\}$ and for $Z\in\{V,U\}$, the decision function $\boldsymbol{f}^\textit{\tiny{Z}}(T_n,\cdot)$, is approximated by a fully connected neural network of the form $\boldsymbol{f}^{\theta_n}\colon\mathbbm{R}^d\to\{0,\,1\}^J$, where $\theta_n\in\mathbbm{R}^{q_n}$ is a vector containing all the $q_n\in\N$ trainable parameters in network $n$. The decision function $\boldsymbol{f}^\textit{\tiny{A}}(T_n,\cdot,\cdot)$ is approximated by similar neural networks, with the only difference that the input also includes information of which derivatives in the portfolio have been exercised prior to $T_n$, \textit{i.e.,} $\boldsymbol{f}^{\theta_n}\colon\mathbbm{R}^d\times\{0,1\}^J\to\{0,\,1\}^J$.

Since binary decision functions are discontinuous, and therefore unsuitable for gradient-type optimization algorithms, we use as an intermediate step, the neural network $\boldsymbol{F}^{\theta_n}\colon\mathbbm{R}^d\to(0,\,1)^J$. Instead of a binary decision, the output of the neural network $\boldsymbol{F}^{\theta_n}$ can be viewed as the probability\footnote{ However the interpretation as a probability may be helpful, one should be careful since it is not a rigorous mathematical statement. It should be clear that there is nothing random about the stopping decisions, since the stopping time is $\Ht-$measurable. It can also be interpreted as a measure on how certain we can be that exercise is optimal.} for exercise to be optimal. This output is then mapped to 1 for values above (or equal to) 0.5, and to 0 otherwise, by defining $\boldsymbol{f}^{\theta_n}(\cdot)=\boldsymbol{\mathfrak{a}}\circ \boldsymbol{F}^{\theta_n}(\cdot)$, where $\boldsymbol{\mathfrak{a}}$ is a component-wise round-off function, \textit{i.e.,} for $j\in\{1,2,\ldots,J\}$, and $x\in\R^d$, the $j$:th component of $\boldsymbol{\mathfrak{a}}(x)$ is given by $\left(\boldsymbol{\mathfrak{a}}(x)\right)_j=\mathbbm{I}_{\{x_j\geq1/2\}}$. For each $Z\in\{V,U\}$, our aim is to adjust the parameters $\theta_1,\theta_2,\ldots,\theta_N$ such that  \begin{align}\label{NN_Z} (\boldsymbol{f}^\textit{\tiny{Z}}(T_1,\cdot),\boldsymbol{f}^\textit{\tiny{Z}}(T_2,\cdot),\ldots,\boldsymbol{f}^\textit{\tiny{Z}}(T_{N},\cdot))
^T\approx
    (\boldsymbol{f}^{\theta_{1}},\boldsymbol{f}^{\theta_{2}},\ldots,\boldsymbol{f}^{\theta_{N}})^T\eqqcolon\boldsymbol{\mathbbm{f}}^{\Theta},\\
    (\boldsymbol{f}^\textit{\tiny{A}}(T_1,\cdot,\cdot),\boldsymbol{f}^\textit{\tiny{A}}(T_2,\cdot,\cdot),\ldots,\boldsymbol{f}^\textit{\tiny{A}}(T_{N},\cdot,\cdot)),
^T\approx
    (\boldsymbol{f}^{\theta_{1}},\boldsymbol{f}^{\theta_{2}},\ldots,\boldsymbol{f}^{\theta_{N}})^T\eqqcolon\boldsymbol{\mathbbm{f}}^{\Theta},\label{NN_A}
    \end{align}
where we recall that $\Theta=\{\theta_1,\theta_2,\ldots,\theta_N\}$. For $n\in\{1,2,\ldots,N\}$, we define the sequence of neural networks, approximating the decision functions at exercise dates $T_n,T_{n+1},\ldots T_N$, by \\$\boldsymbol{\mathbbm{f}}_n^{\Theta}\coloneqq(\boldsymbol{f}^{\theta_{n}},\boldsymbol{f}^{\theta_{n+1}},\ldots,\boldsymbol{f}^{\theta_{N}})^T$. Note that the input dimension for the neural networks is different when we want to approximate $\boldsymbol{f}^\textit{\tiny{V}}$ and $\boldsymbol{f}^\textit{\tiny{U}}$ compared to when we want to approximate $\boldsymbol{f}^\textit{\tiny{A}}$. To avoid having to introduce an extra layer of notation, we use for all networks $\Theta$ to denote the set of parameters, and keep in mind that the dimension depends on the specific problem considered. 
Although the above provides a good intuition for what we want to accomplish, it is not clear in which sense we want the functions to be similar, or how to adjust the parameters to achieve this.  To approach a more tractable form, from a computational perspective, for $t\in(T_{n-1},T_n]$, we insert \eqref{NN_Z} in \eqref{taufstar} and \eqref{NN_A} in \eqref{taufstar_net} to obtain 
\begin{align}\label{tau_Z}
    \boldsymbol{\tau}[\boldsymbol{\mathbbm{f}}^{\Theta}_n](X^{t,x})&=\sum_{k=n}^{N}T_k\boldsymbol{f}^{\theta_k}(X_k)\odot\prod_{m=k}^N\left(\boldsymbol{1}_J- \boldsymbol{f}^{\theta_m}(X_m)\right),\\
    \boldsymbol{\tau}[\boldsymbol{\mathbbm{f}}^{\Theta}_n](X^{t,x},A^{t,\alpha})&=\sum_{k=n}^{N}T_k\boldsymbol{f}^{\theta_k}(X_k,A_k)\odot\prod_{m=k}^N\left(\boldsymbol{1}_J- \boldsymbol{f}^{\theta_m}(X_m,A_m)\right).\label{tau_A}
\end{align}
Note that \eqref{tau_Z} is a $J-$dimensional vector of $X-$stopping times and \eqref{tau_A} is a $J-$dimensional $(X,A)-$stopping times vector, which depends on $\boldsymbol{\mathbbm{f}}^{\Theta}_n$ on a structural level but also on the randomness of the stochastic process $X^{t,x}$ (and $A^{t,\alpha}$ for \eqref{tau_A}). For notational convenience, we use the short hand notation $\bar{\boldsymbol{\tau}}^{\Theta}_n=\boldsymbol{\tau}[\boldsymbol{\mathbbm{f}}^{\Theta}_n](X^{t,x})$ (or $\hat{\boldsymbol{\tau}}^{\Theta}_n=\boldsymbol{\tau}[\boldsymbol{\mathbbm{f}}^{\Theta}_n](X^{t,x},A^{t,\alpha})$, when approximating $\boldsymbol{f}^\textit{\tiny{A}}$), and for element $j\in\{1,2,\ldots,J\}$, $\bar{\tau}
_{n,j}^{\Theta}=\big(\bar{\boldsymbol{\tau}}^{\Theta}_n\big)_j$ (or $\hat{\tau}
_{n,j}^{\Theta}=\big(\hat{\boldsymbol{\tau}}^{\Theta}_n\big)_j$). We are now ready to define our objective, which, for $T_n\in\mathbbm{T}^{\Pi}(0)$, is to find $\theta_n$ such that the expected future cash-flows are maximized. The cash-flows can be divided into three categories: \begin{enumerate}
    \item The cash-flows obtained by the derivatives exercised at the present time $T_n$;
    \item The cash-flows obtained at later exercise dates prior to default of the counterparty;
    \item The cash-flows obtained at default of the counterparty, according to the close-out agreement.
\end{enumerate}
For $n\in\{1,2,\ldots,N\}$ and $j\in\{1,2,\ldots,J\}$, we denote dimension $j$ of decision function $\boldsymbol{f}^{\theta_n}$ by \begin{equation*}
    \big(\boldsymbol{f}^{\theta_n}\big)_j=f_j^{\theta_n},\quad\text{and } \big(\boldsymbol{F}^{\theta_n}\big)_j=F_j^{\theta_n}
\end{equation*}
Given that no default has occurred prior to $T_n$, (and the exercise state $A_n=\alpha$ for the risky portfolio with netting) the expected cash-flows, that we want to maximize, are given below.
\begin{align}\begin{split}\label{theoretical_train_V}
    \textbf{Risk}&\textbf{-free portfolio:}\\\E_{T_n}\Bigg[&\sum_{j=1}^Jf^{\theta_n}_j(X_n)\,g_j(T_n,X_n)+\big(1-f^{\theta_n}_j(X_n)\big)D_{T_{n},\bar{\tau}_{n+1,j}^{V}}g_j\Big(\bar{\tau}_{n+1,j}^\textit{\tiny{V}},X_{\bar{\tau}_{n+1,j}^\textit{\tiny{V}}}\Big)\Bigg],\end{split}\\
    \begin{split}\label{theoretical_train_U}
    \textbf{Risk}&\textbf{y portfolio without netting:}\\
    \E_{T_n}\Bigg[&\sum_{j=1}^Jf^{\theta_n}_j(X_n)\,g_j(T_n,X_n)+\big(1-f^{\theta_n}(X_n)\big)\Big(\mathbbm{1}_{\bar{\tau}_{n+1,j}^\textit{\tiny{U}}}^\text{\tiny{D}}D_{T_{n},\bar{\tau}_{n+1,j}^{U}}g_j\Big(\bar{\tau}_{n+1,j}^{U},X_{\bar{\tau}_{n+1,j}^{U}}\Big)\\
    &+\Big(1-\mathbbm{1}_{\bar{\tau}_{n+1,j}^\textit{\tiny{U}}}^\text{\tiny{D}}\Big)D_{t,\tau^\text{\tiny{D}}}\big(RV_j(\tau^\text{\tiny{D}},X_{\tau^\text{\tiny{D}}})^++V_j(\tau^\text{\tiny{D}},X_{\tau^\text{\tiny{D}}})^-\big)\Big)\Bigg],\end{split}\\\begin{split}\label{theoretical_train_A}
    \textbf{Risk}&\textbf{y portfolio with netting:}\\
    \E_{T_n}\Bigg[&\sum_{j=1}^J\alpha_jf^{\theta_n}_j(X_n,\alpha)\,g_j(T_n,X_n)+\alpha_j\big(1-f^{\theta_n}_j(X_n,\alpha)\big)\mathbbm{1}_{\hat{\tau}_{n+1,j}^\textit{\tiny{U}}}^\text{\tiny{D}}D_{T_{n},\hat{\tau}_{n+1,j}^{U}}g_j\Big(\hat{\tau}_{n+1,j}^{U},X_{\hat{\tau}_{n+1,j}^{U}}\Big)\\
    &+R\bigg(\sum_{k=1}^J\alpha_k\big(1-f^{\theta_n}_k(X_n,\alpha)\big)\Big(1-\mathbbm{1}_{\hat{\tau}_{n+1,k}^\textit{\tiny{U}}}^\text{\tiny{D}}\big)D_{t,\tau^\text{\tiny{D}}}V_k(\tau^\text{\tiny{D}},X_{\tau^\text{\tiny{D}}})\bigg)^+\\
    &+\bigg(\sum_{\ell=1}^J\alpha_\ell\big(1-f^{\theta_n}_\ell(X_n,\alpha)\big)\Big(1-\mathbbm{1}_{\hat{\tau}_{n+1,\ell}^\textit{\tiny{U}}}^\text{\tiny{D}}\Big)D_{t,\tau^\text{\tiny{D}}}V_\ell(\tau^\text{\tiny{D}},X_{\tau^\text{\tiny{D}}})\bigg)^-\Bigg].\end{split}
    \end{align}
    
\begin{equation*}
\end{equation*}
We want to optimize $\theta_n$, such that the above are as close as possible (in mean squared sense) to $\Pi^\textit{\tiny{V}}(T_n,X_n)$, $\Pi^\textit{\tiny{U}}(T_n,X_n,1)$ and $\Pi^\textit{\tiny{A}}(T_n,X_n,1,\alpha)$, respectively.

\begin{remark}\label{V_approx}
The objectives for the risky portfolios, in \eqref{theoretical_train_U} and \eqref{theoretical_train_A}, both depend on the risk-free valuation of the derivatives. Therefore, in order to approximate the risky decision functions, we first need to approximate the risk-free exercise strategy, and the risk-free derivative values. In the next subsection, we explain how $V_j$ can be approximated.
\end{remark}
Although \eqref{theoretical_train_V}-\eqref{theoretical_train_A} are accurate representations of the optimization problems, they give us some practical problems. In general, we have no access to $\boldsymbol{f}^\textit{\tiny{V}}$, $\boldsymbol{f}^\textit{\tiny{U}}$ and $\boldsymbol{f}^\textit{\tiny{A}}$ which control $\bar{\boldsymbol{\tau}}_{n+1}^\textit{\tiny{V}}$, $\bar{\boldsymbol{\tau}}_{n+1}^\textit{\tiny{U}}$ and $\hat{\boldsymbol{\tau}}_{n+1}^\textit{\tiny{A}}$. Another problem is that, in general, we have no access to the true distributions of the portfolio values for comparison. However, if $T_{\tilde{N}}$ is the maturity of derivative $j\in\{1,2,\ldots,J\}$, it is optimal to exercise as long as the pay-off value is positive, by the definition of the decision functions. We can therefore set \begin{equation*}
    f_j^{\theta_{\tilde{N}}}(\,\cdot\,) \coloneqq \I_{\{g_j(T_{\tilde{N}},\,\cdot\,)>0\}}, \quad\text{and}\quad f_j^{\theta_k}(\,\cdot\,) \coloneqq 0, \ \text{for } k>{\tilde{N}}.
\end{equation*}
Furthermore, at $T_N$, the maturity of the portfolio, the positive part of the pay-off value equals the derivative value (if no default in $(T_{N-1},T_N]$, in the risky cases). At $T_{N-1}$, Equation \eqref{theoretical_train_V} then becomes 
\begin{equation}\label{last_step_theoretical}
    \E_{T_{N-1}}\Big[\sum_{j=1}^Jf^{\theta_{N-1}}_j(X_N)\,g_j(T_{N-1},X_{N-1})+D_{N-1,N}\big(1-f^{\theta_{N-1}}_j(X_{N-1})\big)\,g_j(T_N,X_N)\Big].\end{equation}
Recall that if $T_N$ is greater than the maturity of contract $j$, we have $g_j(T_N,\cdot)\equiv0$. Since all components in \eqref{last_step_theoretical} are known except for the decision function $\boldsymbol{f}^{\theta_{N-1}}$, we want to find $\theta_{N-1}$, such that a Monte-Carlo approximations of \eqref{last_step_theoretical} is maximized. Given $M\in\N$ samples, distributed as $X$, which for $m\in\{1,2,\ldots,M\}$ is denoted by $x=(x_{t}(m))_{t\in[0,T]}$, we approximate \eqref{last_step_theoretical} by
\begin{equation}\label{last_step_empirical}
   \frac{1}{M}\sum_{m=1}^{M}\sum_{j=1}^Jf^{\theta_{N-1}}_j(x_{N-1}(m))\,g_j(T_{N-1},x_{N-1}(m))
    +D_{N-1,N}\big(1-f^{\theta_{N-1}}_j(x_{N-1}(m))\big)\,g_j\left(T_N,x_N(m)\right).\end{equation}
The only unknown entity in \eqref{last_step_empirical} is the parameter $\theta_{N-1}$ in the decision function \\$\boldsymbol{f}^{\theta_{N-1}}=(f^{\theta_{N-1}}_1,\ldots,f^{\theta_{N-1}}_j)^T$. Furthermore, we wish to find $\theta_{N-1}$ such that \eqref{last_step_empirical} is maximized, since it represents the average cash-flow in $[t_{N-1},t_N]$. Once $\theta_{N-1}$ is optimized, we use this parameter to set up a similar expression for the expected cash-flow on $[t_{N-2},t_{N}]$, which is maximized by finding an optimal $\theta_{N-2}$. This procedure is then iteratively continued until also $\theta_{N-3}, \theta_{N-4}, \ldots,\theta_{1}$ are optimized. The procedure is similar for the risky portfolios, but based on \eqref{theoretical_train_U} or \eqref{theoretical_train_A} instead. This implies that we also need to sample default events of the counterparty. We denote by $\theta_n^*$ the optimized version of parameter $\theta_n$ and the sequence of optimized parameters for the networks at exercise dates $T_n,T_{n+1},\ldots,T_N$ are defined as \begin{equation*}
    \Theta^*_n\coloneqq\{\theta_n^*,\theta^*_{n+1},\ldots,\theta_N^*\},
\end{equation*}
and for notational convenience, we define the complete sequence of parameters as $\Theta^*\coloneqq\Theta_1^*$.
\begin{remark}
Since we are considering a portfolio in which all the derivatives may have a different set of exercise dates, we have that for $T_n\in\mathbbm{T}^\Pi(0)$, there are $J_n^{\emph{Ex}}\in\{1,2,\ldots,J\}$ derivatives that may be exercised. Therefore, we only need to compute $J_n^{\emph{Ex}}$ of the $J$ dimensions of $\boldsymbol{f}^{\theta_n}$ and can by default set the remaining $J-J_n^{\emph{Ex}}$ dimensions of $\boldsymbol{f}^{\theta_n}$ to 0. This can be done by appropriately adjusting some weights and biases. 
\end{remark}
To keep the flow of the paper, the details of the algorithms and the parameters $\theta_n$ are given in the Appendix.

\subsection{Phase II: Learning pathwise derivative values and portfolio exposures}
As mentioned in Remark \ref{V_approx}, the risk-free derivative values need to be approximated pathwise in order to approximate the risky decision functions. Moreover, the pathwise derivative values are required to approximate the exposure profiles for the risk-free, as well as the risky portfolios. In this subsection, we focus on the risk-free portfolios, but the extension to risky portfolios is straight-forward.

We use the risk-free, stopping strategy from subsection \ref{DOSalgo} to generate pathwise cash-flows, which are in turn used to approximate the pathwise derivative values. Let $T_{n-1},T_n\in\mathbbm{T}^{\Pi}(0)$, for $t\in(T_{n-1},T_n]$, we denote the vector-valued discounting process and pay-off function, respectively, by \begin{equation}
\boldsymbol{D}(t,\boldsymbol{\tau}_n^\textit{\tiny{V}})\coloneqq\begin{pmatrix}D(t,\tau_{n,1}^\textit{\tiny{V}})\\
    \vdots\\
    D(t,\tau_{n,J}^\textit{\tiny{V}})
    \end{pmatrix}
    \quad \text{and,} \quad 
    \boldsymbol{g}(\boldsymbol{\tau}_n^\textit{\tiny{V}}, X^{t,x})\coloneqq\begin{pmatrix}g_1(\tau_{n,1}^\textit{\tiny{V}},X_{\tau_{n,1}^\textit{\tiny{V}}})\\
    \vdots\\
    g_J(\tau_{n,J}^\textit{\tiny{V}},X_{\tau_{n,J}^\textit{\tiny{V}}})
    \end{pmatrix}.
\end{equation}
Using the notation above, we define the vector-valued cash-flow process as
\begin{equation}
    \boldsymbol{Y}_t\coloneqq\boldsymbol{D}_{t,\boldsymbol{\tau}^\textit{\tiny{V}}_n}\odot\boldsymbol{g}(\boldsymbol{\tau}^\textit{\tiny{V}}_n, X_{\boldsymbol{\tau}^\textit{\tiny{V}}_n}),
\end{equation}
and for $j\in\{1,2,\ldots,J\}$, we denote the $j$:th element of $\boldsymbol{Y}_{t}$ by $Y_{t,j}$, and we emphasize that $\boldsymbol{Y}_t$ is not $\Ht-$measurable. 
\subsubsection{Regression problems}
In this subsection we use standard regression theory, see \textit{e.g.,} \cite{regression_book}, to show that derivative values, exposures, and other entities of interest, can be formulated as the solution to certain minimization problems. We introduce the following notation for measurable functions\footnote{We assume measurable spaces $(C_1,\mathcal{C}_1)$ and $(C_2,\mathcal{C}_2)$ and measurable functions with respect to $\sigma$-algebras $\mathcal{C}_1$ and $\mathcal{C}_2$. This assumption holds for all cases in this paper.}
\begin{equation}\label{measurable}
    \D(C_1;C_2)\coloneqq\{\,f\colon C_1\to C_2\,|\,f\ \text{measurable}\}. 
\end{equation}
First, we recall a basic property of the regression function. Let $\mathcal{X}\colon[0,T]\times\Omega\to C_1$ and $\mathcal{Y}\colon[0,T]\times\Omega\to C_2$ be a stochastic process, which for $t,u\in[0,T]$, with $t\leq u$, satisfies $\E_0[|\mathcal{Y}_t|^2]<\infty$. We define the regression function, which satisfies \begin{equation}\label{reg_fun}
    m(t,\cdot)\in\argmin_{\boldsymbol{h}\in\mathcal{D}(C_1;C_2)}\E_{t}\left[\left\|\boldsymbol{h}(\mathcal{X}_t)-\mathcal{Y}_u\right]\right\|_2^2],
\end{equation}
where $\|\cdot\|_2$ is the Euclidean norm. It then holds, for $x\in\R^a$, that the regression function is given by the conditional expectation \begin{equation}\label{reg_det}
    m(t,\chi)=\E^\Q\big[\mathcal{Y}_u\,\big|\,\mathcal{X}_t=\chi\big].
\end{equation}
Using the notation from above, and by choosing $C_1, C_2$, $\mathcal{X}$ and $\mathcal{Y}$ in \eqref{reg_fun} and \eqref{reg_det} wisely, we can approximate different entities related to \textit{e.g.,} the exposure profiles or the pathwise CVA. For instance the exposures, both with and without netting, can be approximated from the solution of a minimization problem of the form in \eqref{reg_fun}. For $T_{n-1},T_n\in\mathbbm{T}^\Pi(0)$, let $t\in(T_{n-1},T_n]$ and denote $\boldsymbol{V}(t,\,\cdot\,)=\left(V_1(t,\,\cdot\,),\ldots V_J(t,\,\cdot\,)\right)^T$, it then holds that
\begin{align}\label{reg1}
    \boldsymbol{V}(t,\cdot)\in&\argmin_{\boldsymbol{h}\in\mathcal{D}(\R^d;\R^J)}\E_t\left[\left\|\boldsymbol{h}(X_t)-\boldsymbol{Y}_t\right\|_{2}^2\right],\\
     \sum_{j=1}^JV_j(t,\cdot)(A_t)_j\in&\argmin_{h\in\mathcal{D}(\R^d\times\{0,1\}^J;\R)}\E_t\bigg[\Big|h(X_t,A_t)-\sum_{j=1}^JY_{t,j}(A_t)_j\Big|^2\bigg].\label{reg2}
\end{align}
In \eqref{reg1}, we have $C_1=\R^d$, $C_2=\R^J$, $\mathcal{X}=X$ and $\mathcal{Y}=\boldsymbol{Y}$ and in \eqref{reg2}, we have $C_1=\R^d\times\{0,1\}^J$, $C_2=\R$, $\mathcal{X}=(X,A)$ and $\mathcal{Y}=\sum_{j=1}^JY_{\cdot,j}(A)_j$.  For $s\in[0,T]$ and for $j\in\{1,2,\ldots,J\}$, by \eqref{assumtion2int}, it holds that $\E_0\left[|Y_{s,j}|^2\right]<\infty$, and therefore also $\E_0\left[\|\boldsymbol{Y}_{s}\|_2^2\right]<\infty$. Now, \eqref{reg1} holds trivially since by definition $ \boldsymbol{V}(t,x)=\boldsymbol{\E}_{t,x}\left[\boldsymbol{D}_{t,\boldsymbol{\tau}^\textit{\tiny{V}}_n}\odot\boldsymbol{g}(\boldsymbol{\tau}^\textit{\tiny{V}}_n, X_{\boldsymbol{\tau}^\textit{\tiny{V}}_n})\right]=\boldsymbol{\E}_{t,x}\left[\boldsymbol{Y}_t\right]$. For \eqref{reg2}, it follows that 
\begin{align}
    \sum_{j=1}^JV_j(t,x)(A_t)_j&=\sum_{j=1}^J\E_{t,x}\left[D_{t,\tau_{n,j}^\textit{\tiny{V}}}g_j(\tau_{n,j}^\textit{\tiny{V}},X_{\tau_{n,j}^\textit{\tiny{V}}})\right](A_t)_j\\
    &=\E_{t,x}\left[\sum_{j=1}^J D_{t,\tau_{n,j}^\textit{\tiny{V}}}g_j(\tau_{n,j}^\textit{\tiny{V}},X_{\tau_{n,j}^\textit{\tiny{V}}})(A_t)_j\right]=\E_{t,x}\left[\sum_{j=1}^J Y_{t,j}(A_t)_j\right],
\end{align}
where we have used linearity of expectations and the fact that $A_t$ is $\Ht-$measurable.

\subsection{Neural network based regression algorithm}
Since the specific details of the neural networks are transferable from Appendix \ref{NN_spec} and \ref{train_NN}, this subsection is less detailed. The main idea is to represent $\mathcal{D}(\R^\textit{\tiny{A}};\R^b)$ (measurable functions from $\R^\textit{\tiny{A}}$ to $\R^b$, defined in \eqref{measurable}) by a parametrized neural network. A minimization problem of the form \eqref{reg_fun} can then be used as a loss function, which should be minimized by adjusting some set of trainable parameters. However, in general we have no access to $\boldsymbol{\tau}_n^\textit{\tiny{V}}$ for $n<N$, where $N=|\mathbbm{T}
^\Pi(0)|$. On the other hand, we can use the exercise strategy from Subsection \ref{DOSalgo}, \textit{i.e.,} approximate $\boldsymbol{\tau}_n^\textit{\tiny{V}}$ by $\bar{\boldsymbol{\tau}}^{\Theta^*}_n=\big(\bar{\tau}_{n,1}^{\Theta^*},\ldots,\bar{\tau}_{n,J}^{\Theta^*}\big)^T$. Furthermore, $\E_t[\cdot]$ in \eqref{reg_det} needs to be approximated by Monte-Carlo samples. We use $M_\text{reg}\in\N$ samples, distributed as $X$ and $\boldsymbol{Y}$, which for $m\in\{1,2,\ldots,M_\text{reg}\}$ are denoted by\footnote{ In practice we set $x_\text{reg}=x_\text{val}$, where $x_\text{val}$ is defined in Phase I.} $x
^\text{reg}(m)=(x^\text{reg}_{t}(m))_{t\in[0,T]}$ and $\boldsymbol{y}(m)=(\boldsymbol{y}_{t}(m))_{t\in[0,T]}$. Furthermore, we use \begin{equation*}A_t(m)\approx A^{\Theta^*_n}_t(m)=\begin{pmatrix}\I_{\big\{\bar{\tau}
_{n,1}^{\Theta^*}(m)>t\big\}},\\
\vdots\\
\I_{\big\{\bar{\tau}
_{n,J}^{\Theta^*}(m)>t\big\}}
\end{pmatrix}\end{equation*}
where for $j\in\{1,2,\ldots,J\}$, $\bar{\tau}
_{n,j}^{\Theta^*}(m)=\left(\boldsymbol{\tau}[\boldsymbol{\mathbbm{f}}^{\Theta^*}_n](x^{t,x^\text{reg}_t(m)}(m))\right)_j$.

We define for $n\in\{1,2,\ldots,N\}$, the neural networks $\boldsymbol{h}^{\Phi_n}\colon\R^d\to\R^J$ and $h^{\Phi_n}\colon\R^d\times\{0,1\}^J\to\R$, which are parametrized by $\Phi^\text{\tiny{IR}}_n\in\R^{u^\text{\tiny{IR}}_n}$ and $\Phi^\text{\tiny{PR}}_n\in\R^{u^\text{\tiny{PR}}_n}$ where $u_n^\text{\tiny{IR}},u_n^\text{\tiny{PR}}\in\N$ are the number of trainable parameters in each network. We use as loss functions, the empirical counterparts of \eqref{reg1} and \eqref{reg2}, which are given by \begin{align}\label{DOS-IR}
   \text{DOS-IR:}\quad\quad &\frac{1}{M_\text{reg}}\sum_{m=1}^{M_\text{reg}}\|\boldsymbol{h}^{\Phi_n^1}(x^\text{reg}_t(m))-\boldsymbol{y}_t(m)\|_2^2,\\
    \text{DOS-PR:} \quad\quad &\frac{1}{M_\text{reg}}\sum_{m=1}^{M_\text{reg}}\Big|h^{\Phi_n^2}\big(x^\text{reg}_t(m),A^{\Theta^*_n}_t(m)\big)-\sum_{j=1}^J y_{t,j}(m)\big(A^{\Theta^*_n}_t(m)\big)_j\Big|^2.\label{DOS-PR}
\end{align}
''DOS'' in DOS-IR and DOS-PR refers to the fact that the deep stopping strategy used to obtain $\boldsymbol{y}(m)$ is generated by the DOS-algorithm. 'IR' and 'PR' are abbreviations for ''individual regression'' and ''portfolio regression'', and refer to the regression at the level of each individual derivative, and the regression at portfolio level given in \eqref{DOS-IR} and in \eqref{DOS-PR}, respectively. 

The exact algorithm for computations of pathwise CVA in order to obtain ES-CVA is not presented in details here. However, it is straight-forward to adjust \eqref{DOS-IR} and \eqref{DOS-PR}, to approximate pathwise CVA instead. 

The only important adjustment to the structure of the neural networks (details in Appendix \ref{NN_spec}) is that we want the output to be unbounded and therefore use the identity as scalar activation function in the output layers.

\subsection{Combining Phase I and Phase II}
Recall that the purpose for using regression at the level of each derivative was to be able to approximate the exposure of a portfolio of derivatives without netting agreement. If we consider derivatives with non-negative value, the definitions of exposures with and without netting agreements coincide. Therefore, only derivatives with non-negative values are considered in this paper, to be able to compare regression on derivative level with the regression on portfolio level. For $T_{n-1},T_n\in\mathbbm{T}^\Pi(0)$, let $t\in(T_{n-1},T_n]$, we define the following approximators
\begin{align}\label{V_DOS-IR}
    \boldsymbol{V}^{\text{DOS-IR}}\left(t,\,\cdot\, \big|\,\Phi_n^\text{\tiny{IR}},\Theta^*\right)&\coloneqq\boldsymbol{h}^{\Phi^\text{\tiny{IR}}_n}(\cdot),\quad \text{(individual derivative values)},\\\label{cw-EE}
      \textbf{E}^{\text{DOS-IR}}\left(t,\,\cdot\, \big|\,\Phi_n^\text{\tiny{IR}},\Theta^*\right)&\coloneqq\boldsymbol{h}^{\Phi^\text{\tiny{IR}}_n}(\cdot)\odot A_t^{\Theta^*},  \quad\text{(derivative exposures)},\\\label{Pi_DOS-IR}
    \text{E}^{\text{DOS-IR}}\left(t,\,\cdot\, \big|\,\Phi_n^\text{\tiny{IR}},\Theta^*\right)&\coloneqq\sum_{j=1}^J\big(\boldsymbol{h}^{\Phi^\text{\tiny{IR}}_n}(\cdot)\big)_j(A^{\Theta^*}_t)_j, \quad\text{(portfolio exposure)},\\\label{PI_DOS-PR}
    \text{E}^{\text{DOS-PR}}\left(t,\,\cdot\, \big|\,\Phi_n^\text{\tiny{PR}},\Theta^*\right)&\coloneqq h^{\Phi_n^\text{\tiny{PR}}}(\cdot,A^{\Theta^*}_t), \quad\text{(portfolio exposure)},
\end{align}
where $\Phi_n^\text{\tiny{IR}}$, and $\Phi_n^\text{\tiny{PR}}$ are parameters optimized by minimizing \eqref{DOS-IR} and \eqref{DOS-PR}, respectively, and $\Theta^*$ are parameters optimized according to the procedure described in the training procedure, described in Phase I, and $\boldsymbol{\mathbbm{I}}_t\in\{0,1\}^J$ represents the exercise history of each derivative in the portfolio. Note that, even though $\Theta^*$ does not appear explicitly in the right hand side of \eqref{V_DOS-IR}, it is crucial since the cash-flow vector $\boldsymbol{y}$, used in \eqref{DOS-IR} and \eqref{DOS-PR}, is created by applying an exercise strategy controlled by $\Theta^*$.
The approximations of EE and PFE are constructed from $M_{\text{train}}\in\mathbbm{N}$ independent realizations of $X$, which for $m\in\{1,\,2,\,\ldots,\,M\}$ are denoted by  $(x_t^{\text{train}}(m))_{t\in[0,T]}$. For $z\in\{\text{IR},\text{PR}\}$, the approximators are given by
\begin{align}\label{EE_DOS-z}
    \widehat{\text{EE}}^{\text{DOS-}z}(t)&\coloneqq\sum_{m=1}^M \Pi^{\text{DOS-}z}\left(t,x(m)\, \big|\,\Phi_n^{z},\Theta^*\right),\\
    \widehat{\text{PFE}}_\alpha^{\text{DOS-}z}(t)&\coloneqq\Pi^{\text{DOS-}z}\left(t,x(i_\alpha)\, \big|\,\Phi_n^{z},\Theta^*\right),\label{PFE_DOS-z}
\end{align}
where $i_\alpha$ is the index of the empirical $\alpha-$percentile of the vector \\$\left(\Pi^{\text{DOS-}z}\left(t,x(1)\, \big|\,\Phi_n^{z},\Theta^*\right),\ldots,\Pi^{\text{DOS-}z}\left(t,x(M)\, \big|\,\Phi_n^{z},\Theta^*\right)\right)$.

\section{Numerical experiments}\label{sec4}
In the numerical experiments we use a Geometric Brownian Motion (GBM) to model the asset processes and an intensity model for default events of the counterparty. To be able to incorporate WWR, the default intensity is linked to the market state of the asset processes. The default event is triggered by an exogenous component, independent of observable market information. On the other hand, the intensity depends on the credit spread of the counterparty (observable from zero-coupon bonds) as well as a WWR-parameter. Our model choices are not necessarily used in practice but they serve the purpose of being easy to analyse. Especially the default model makes it straight-forward to analyze the effects of the credit spread of the counterparty and the WWR-parmeter. It should be pointed out that the algorithms described in this paper are model independent in the sense that they are fully data driven. This means that as long as we can sample (or in any other way obtain) market data and default events, we can train the neural networks and the computations below can be performed.

In addition, the algorithms have also been implemented for a portfolio of Bermudan swaptions with dynamics following the one-factor Hull--White model. The results are similar, and are therefore not included in this section.

\subsection{Risk-factor model}
In the Black--Scholes framework, the assets are described by a $\N_+\ni d-$dimensional Geometric Brownian. For $t\in[0,T]$, with constant risk-free rate $r\in\R$, initial state $s_0,\in(0,\infty)^d$, constant dividend $q\in(0,\infty)^d$ and volatility $\sigma\in(0,\infty)^d$, component $i\in\{1,2,\ldots,d\}$ of the asset process $S=(S_t)_{t\in[0,T]}$ is given by
\begin{equation}\label{S}
    (S_t)_i=(s_0)_i\,\text{exp}\Big({\Big(r-q_i-\frac{\sigma_{i}^2}{2}\Big)t + \sigma_{i}(W_t)_i}\Big),
\end{equation}
where $W=(W_t)_{t\in[0,T]}$ is a correlated standard Brownian motion, satisfying for $i,j\in\{1,2,\ldots,d\}$, \\$\E_0[\d(W_t)_i\d(W_t)_j]=\rho_{ij}\d t$, with $\rho_{ij}\in[-1,1]$. 
\subsection{Default model}
Following \cite{SGBM_WWR}, we model a default event of the counterparty as \begin{equation*}\tau^\text{\tiny{D}}=\inf_{t\in[0,T]}\Big\{t\colon\int_0^t\tilde{h}(u,\tilde{S}_u)\d u\geq E_1\Big\},\end{equation*} where $E_1$ is a random variable, uniformely distributed on $[0,1]$. Furthermore, the process $(\tilde{S}_t)_{t\in[0,T]}$ is the geometric average of the $d$ component of the dividend-free version of $S$, given by
\begin{equation*}
    \tilde{S}_t=\prod_{i=1}^d\bigg((s_0)_i\,\text{exp}\Big({\Big(r-\frac{\sigma_{i}^2}{2}\Big)t + \sigma_{i}(W_t)_i}\Big)\bigg)^{1/d}.
\end{equation*}
For simplicity, without loss of generality, from now on, we assume no correlation between the components of the Brownian motions, \textit{i.e.,} $\rho_{ij}=0$ for $i\neq j$. The above is a one-dimensional GBM, which can be written as
\begin{equation*}
    \tilde{S}_t=\tilde{s}_0\,\text{exp}\bigg(\Big(\tilde{\mu}-\frac{\tilde{\sigma}^2}{2}\Big)t+\tilde{\sigma}\tilde{W}_t\bigg),
\end{equation*}
where $\tilde{s}_0=\Big(\prod_{i=1}^d(s_0)_i\Big)^{1/d}$, $\tilde{\sigma}=\frac{1}{d}\Big(\sum_{i=1}^d\sigma_{i}^2\Big)^{1/2}$, $\tilde{\mu}=r-\frac{1}{2d}\big(1-\frac{1}{d}\big)\sum_{i=1}^d\sigma_i^2$ and $\tilde{W}_t=\frac{1}{d\tilde{\sigma}}\sum_{i=1}^d\sigma_i(W_t)_i$. For $(t,x)\in[0,T]\times\R_+$, $\tilde{h}$ is of the form $\tilde{h}(t,x)=c(t)+b\log{x}$. It can be checked that $\tilde{W}=(\tilde{W}_t)_{t\in[0,T]}$, is a 1-dimensional standard Brownian motion. By setting \begin{equation*}
    c(t)=\bar{h} + b\log{\tilde{S}_0}-\big(r-\frac{\tilde{\sigma}^2}{2}\big)bt+\frac{1}{2}b^2\tilde{\sigma}^2t^2,\end{equation*} we obtain \begin{equation*}
    \tilde{h}_t=\tilde{h}(t,\tilde{S}_t)=\bar{h}+\frac{1}{2}\tilde{\sigma}^2t^2b^2+b\tilde{\sigma}\tilde{W}_t.
\end{equation*}
The economic interpretation of the above is that $\bar{h}$ is the credit spread for the counterparty and $b$ controls the wrong way risk (WWR). 

Recall that the jump-to-default process, $\mathbbm{1}^\text{\tiny{D}}$, is given by $\mathbbm{1}_t^\text{\tiny{D}}=\I_{\{t<\tau^\text{\tiny{D}}\}}$ which gives a survival probability $G_t=\E^\Q[\mathbbm{1}_t^\text{\tiny{D}}|\mathcal{H}_t]=\text{exp}\Big(-\int_0^t\tilde{h}_s\d s\Big)$ (for details, see \cite{SGBM_WWR}). 

\subsection{Experiments}\label{experiments}
\textbf{Contract details:}\newline
We consider a portfolio of $J=8$ derivatives, depending on an asset process in $d=2$ dimensions. We set $T=3$ and use for each derivative, $j\in\{1,2,\ldots,8\}$, the set of exercise dates $\mathbbm{T}_j(t)=\mathbbm{T}_j=\left\{0       , \frac{1}{3}, \frac{2}{3}, 1       , \frac{4}{3},
       \frac{5}{3}, 2        , \frac{7}{3}, \frac{8}{3}, 3        \right\}$, and the pay--off functions given in Table \ref{pay-offs}.
     
\begin{table}[htp]
\centering
\begin{tabular}{lll}
\textbf{Contract number} & \textbf{Contract name} & \textbf{Pay-off function}   \\ \hline
   \textbf{j=1}       & Max-call option  &  $\left(\max\{x_1,x_2\}-100\right)^+$\\
  \textbf{j=2}      & Max-put option &  $\left(100-\max\{x_1,x_2\}\right)^+$\\
  \textbf{j=3}       & Geometric-average-call option & $\left(\sqrt{x_1x_2}-100\right)^+$ \\
  \textbf{j=4}       & Geometric-average-put option & $\left(100-\sqrt{x_1x_2}\right)^+$ \\
  \textbf{j=5}        & Arithmetic-average-call option & $\left(\frac{1}{2}(x_1+x_2)-100\right)^+$ \\
  \textbf{j=6}        & Arithmetic-average-put option &  $\left(100-\frac{1}{2}(x_1+x_2)\right)^+$\\
  \textbf{j=7}        & 1d-call option & $\left(x_1-100\right)^+$ \\
  \textbf{j=8}        & 1d-put option & $\left(100-x_1\right)^+$
\end{tabular}
 \caption{The two components of the asset process, $S$, are represented by $x_1,x_2\in\R$ and $\left(\,\cdot\,\right)^+=\max\{\,\cdot\,,0\}$.}
  \label{pay-offs}
\end{table} 
\noindent\textbf{Dynamics details:}\\
For $i\in\{1,2\}$, we set $(s_0)_{i}=100$, $q_i=0.1$, $r=0.05$, $\sigma_i=0.2$, $\rho_{ii}=1$ and $\rho_{12}=\rho_{21}=0$.

\noindent\textbf{Neural network details:}\newline
We use $M_\text{train}=M_\text{reg}=M_\text{reg}=M=2^{20}$,
and for simplicity, the same structure and hyperparameter-settings are used in all networks, \textit{i.e.,} all the networks in Phase I, and Phase II. We use training batches of size 5000, 3 hidden layers and 30 nodes in each hidden layer. Furthermore, the learning rate decreases step-wise, with equally sized steps after 100 training batches from $10^{-2}$ to $10^{-6}$.       
\subsubsection{Risk-free valuation}\label{risk-free}
The purpose of the risk-free valuation is two-fold. Firstly, since the risky derivative values are used as an input in the risky-valuations, they need to be accurate. To ensure accuracy, the risk-free values are compared to a well-established existing valuation method, namely the Stochastic Grid Bundling Method (SGBM), see \cite{SGBM} for details. Secondly, it is in its own, an interesting and challenging problem to compute the value of a portfolio of complex derivatives with early-exercise features, without having to do one computation per derivative. 

\begin{figure}[htp]
\centering
\begin{tabular}{ccc}
     \includegraphics[width=80mm]{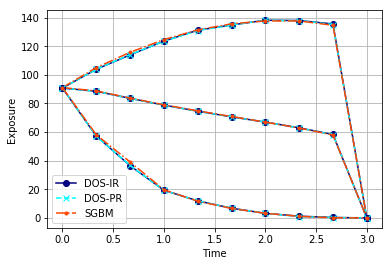}  & \includegraphics[width=80mm]{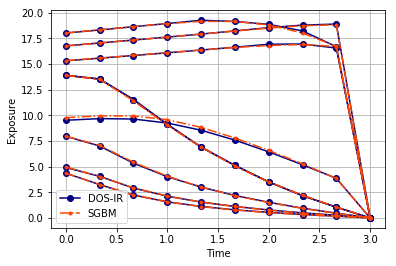}
\end{tabular}
\caption{\textbf{Left:} EE, PFE$_{97.5}$ and PFE$_{2.5}$ at portfolio level computed with DOS-PR, DOS-IR and SGBM respectively. \textbf{Right:} EE at derivative level computed with DOS-IR and SGBM, respectively.}
\label{BS-fig}
\end{figure}

As mentioned above, we compare the algorithms introduced in earlier sections with the SGBM. We emphasize that the values for each derivative needs to be approximated individually when using SGBM, \textit{i.e.,} we perform 8 regressions, one for each derivative. In Table \ref{prices}, we compare the value at $t=0$ for each derivative approximated with the portfolio version of the DOS-algorithm and the SGBM. In Table \ref{prices}, we compare our values for each derivative at the initial time with the values obtained by SGBM. For the DOS-values, the neural network is trained five times, and evaluated on new, independent, samples and the average values are reported. For the SGBM, the regression is performed five times for each derivative, and the average values of the direct estimators are reported. It should be mentioned that, for $j\in\{3,4,5,6,7,8\}$, the SGBM-values are biased high\footnote{The reason that the SGBM-values are not biased high for $j\in\{1,2\}$ can be explained by the non-linearity of the max-operator in the pay-off functions. For more details we refer to \cite{SGBM}.} and, for all $j$, the DOS-values are biased low. 
\begin{table}[htp]\centering
\begin{tabular}{|l|l|l|l|l|l|l|l|l|l|}
\hline
 & $\textbf{j=1}$ & $\textbf{j=2}$ & $\textbf{j=3}$ & $\textbf{j=4}$ & $\textbf{j=5}$ & $\textbf{j=6}$ & $\textbf{j=7}$ & $\textbf{j=8}$ & $\Pi(0,s_0)$ \\ \hline
     \textbf{Ref.}     &   13.902        &    NA      &     NA     &   NA        &  NA         &     NA  & NA & NA & NA   \\ \hline
     \textbf{DOS} &  13.902 & 9.520 & 4.363 & 16.770 & 4.919 & 15.313 & 7.965 & 18.021 & 90.773 \\ \hline
      \textbf{SGBM}    &  13.899 & 9.780 & 4.366 & 16.770 & 4.971 & 15.327 & 7.963 & 18.033 & 91.108   \\ \hline
\end{tabular}
\caption{Valuation at the level of each derivative with the portfolio version of DOS, and the SGBM. The reference solution is computed by a binomial lattice model in \cite{Binomial_price}.}\label{prices}
\end{table}
In Figure \ref{BS-fig} to the left we compare EE, PFE$_{97.5}$ and PFE$_{2.5}$ approximated with SGBM and according to \eqref{EE_DOS-z} and \eqref{PFE_DOS-z}. To the right, we compare the derivative-wise EE approximated with SGBM and the DOS-IR based algorithm. In the DOS-IR based algorithm, the EE is approximated by evaluating \eqref{cw-EE} at $x(1),\ldots,x(M)$ and computing the component-wise sample mean.

We conclude that, although very different in nature, the SGBM and the two methods presented in this paper, agree on the values at time 0, the tail-distribution of the portfolio exposure over time (at least at the 97.5 and 2.5 percentiles), and the average values of each derivative over time. We emphasize, that we do not claim that one of the methods perform better than the others. For an analysis of the difference in performance between the methods (in the special case $J=1$), we refer to \cite{DOSexposure}.  

\subsubsection{Risky valuation}\label{risky}
In this subsection we focus on the impact of the exercise policy for CVA computations. To be more precise, we investigate to what extent the CVA is overestimated when using the risk-free exercise policy, with and without netting, for different levels of WWR and credit quality of the counterparty. The contracts described in Table \ref{pay-offs} all have positive pay-offs meaning that the corresponding derivative values are also positive. This eliminates the netting effect, and therefore, we add a derivative to the portfolio, which may take on negative values. The 9:th derivative is a European-type future with pay-off $g_9(t,S_t)=2\times\big(80-(S_t)_1\big)\I_{\{t=T\}}$. By the martingale property we have that \begin{equation*}
    V_9(t,S_t)=\E_t\big[\text{e}^{-r(T-t)}g_9(T,S_T)\big]=160\text{e}^{-r(T-t)}-2(S_t)_1\text{e}^{-q_1(T-t)}.
\end{equation*}
Furthermore, in the netted portfolio, we include an interest rate-free collateral, $
C=35$. The collateral is such that, at a default, the banks exposure is lowered by 35, and added to the close-out amount. If no default occurs before maturity, then the collateral plays no role. 
\begin{table}[htp]
\centering
\begin{tabular}{|p{1.5cm}|p{1cm}||p{2cm}|p{2cm}|p{2cm}|p{2cm}|p{2cm}|} \hline
    {} & {$\bar{h}$} & {$\Upsilon^\textit{\tiny{V}}\big[\boldsymbol{f}^\textit{\tiny{V}}\big]$} & {$\Upsilon^\textit{\tiny{U}}\big[\boldsymbol{f}^\textit{\tiny{U}}\big]$} & {$\Upsilon^\textit{\tiny{U}}\big[{\boldsymbol{f}}^\textit{\tiny{V}}\big]$} & {$\Upsilon^\textit{\tiny{A}}\big[{\boldsymbol{f}}^\textit{\tiny{A}}\big]$} &{$\Upsilon^\textit{\tiny{A}}\big[{\boldsymbol{f}}^\textit{\tiny{V}}\big]$}  \\ \hline
      & 0 & 78.62 & 77.47 & 77.41 & 77.86 & 77.84 \\
    $b=-0.2$ & 0.1&78.62 & 59.07 & 57.22  & 69.20  & 67.72 \\
     & 0.2  &78.62 & 47.58 & 42.99 & 64.17  & 60.82  \\ \hline
      & 0   &78.62 & 78.62 &  78.62  & 78.62 & 78.62  \\
    $b=0$  & 0.1  &78.62 & 59.50 & 58.00  &  69.55 & 68.39  \\
      & 0.2 &78.62 & 47.83 & 43.58  &   64.37  & 61.32\\ \hline
      & 0  &78.62 & 78.28   & 78.27 & 78.59 & 78.58\\
    $b=0.2$ & 0.1 &78.62 & 59.65   & 58.31 & 69.78 & 68.75  \\
     & 0.2   &78.62 & 47.89 & 43.78  & 64.51 & 61.55 \\
 \hline
\end{tabular}
\caption{Portfolio valuation with and without netting for different early-exercise policies. We use the short hand notations $\Upsilon^\textit{\tiny{V}}\big[\boldsymbol{f}\big]=\Upsilon^\textit{\tiny{V}}\big(0,s_0\,|\,\boldsymbol{f}\big)$, $\Upsilon^\textit{\tiny{U}}\big[\boldsymbol{f}\big]=\Upsilon^\textit{\tiny{U}}\big(0,s_0,1\,|\,\boldsymbol{f}\big)$ and $\Upsilon^\textit{\tiny{A}}\big[\boldsymbol{f}\big]=\Upsilon^\textit{\tiny{A}}\big(0,s_0,1,\boldsymbol{1}_9\,|\,\boldsymbol{f}\big)$. Furthermore, in this table $\boldsymbol{f}^\textit{\tiny{V}}$, $\boldsymbol{f}^\textit{\tiny{U}}$ and $\boldsymbol{f}^\textit{\tiny{A}}$ represent our neural network approximations of the optimal decision functions.}\label{tab_upsilon}
\end{table}

\begin{table}[htp]
\centering
\begin{tabular}{|p{1.5cm}|p{1cm}||p{2cm}|p{2cm}|p{2cm}|} \hline
    {} & {$\bar{h}$} & {$\text{CVA}$} & {$\overline{\text{CVA}}$} & {Rel. ov.est.}  \\ \hline
      & 0 & 1.15 & 1.21 & 5.2\% \\
    $b=-0.2$ & 0.1& 19.55  & 21.40  & 9.5\%  \\
     & 0.2  & 31.04 & 35.63  & 14.8\%  \\ \hline
      & 0   & 0 &  0  & 0\%  \\
    $b=0$  & 0.1  & 19.2 & 20.62  &  7.4\%  \\
      & 0.2 & 30.79 & 35.04  &     13.8\%  \\ \hline
      & 0  & 0.34  & 0.35   & 2.9\%   \\
    $b=0.2$ &0.1& 18.97 & 20.31  &  7.1\%  \\
     & 0.2   & 30.73 & 34.84  & 13.4\% \\
 \hline
\end{tabular}
\caption{CVA for a portfolio without netting. The relative error in the column to the right represents, in percentage, the overestimation of the CVA by using only the risk-free policy for early-exercise. 'Rel. ov.est.' is short for 'Relative overestimation'.}\label{tab_no_net}
\end{table}

\begin{table}[htp]
\centering
\begin{tabular}{|p{1.5cm}|p{1cm}||p{2cm}|p{2cm}|p{2cm}|} \hline
    {} & {$\bar{h}$} & {$\text{CVA}^\text{net}$} & {$\overline{\text{CVA}}^\text{net}$} & {Rel. ov.est.}  \\ \hline
      & 0 & 0.78 & 0.834 & 6.4\% \\
    $b=-0.2$ & 0.1& 9.42  & 10.90  & 15.7\%   \\
     & 0.2  & 14.45 & 17.80  & 23.2\%      \\ \hline
      & 0   & 0 &  0  & 0\%  \\
    $b=0$  & 0.1  & 9.07 & 10.23  & 11.4\%  \\
      & 0.2 & 14.25 & 17.30  &   21.4\%  \\ \hline
      & 0  & 0.0364 & 0.0374  & 5.6\%  \\
    $b=0.2$ &0.1& 8.83   & 9.87   & 11.7\%   \\
     & 0.2   & 14.11 & 17.07   & 21.0\% \\
 \hline
\end{tabular}
\caption{CVA for a portfolio with netting. The relative error in the column to the right represents, in percentage, the overestimation of the CVA by using only the risk-free policy for early-exercise. 'Rel. ov.est.' is short for 'Relative overestimation'.}\label{tab_net}
\end{table}

\begin{figure}[htp]
\centering
\begin{tabular}{ccc}
     \includegraphics[width=80mm]{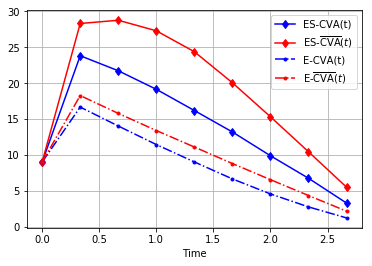}  & \includegraphics[width=80mm]{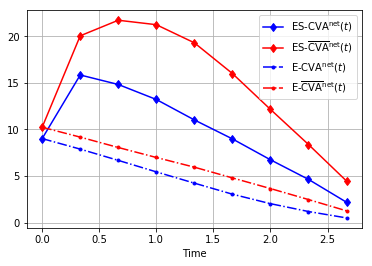}
\end{tabular}
\caption{Time 0 expectation and expected shortfall at level $97.5\%$ of CVA for the risky-free and the risky exercise policies. \textbf{Left:} Without netting. \textbf{Right:} With netting.}
\label{ES-fig}
\end{figure}

 In Table \ref{tab_upsilon}, the portfolio values with and without netting are displayed for different exercise policies and for different credit qualities and WWR-parameters. We see that for low values of $\bar{h}$ (credit spread), \textit{i.e.,} for high credit quality of the counterparty, the risk-free exercise policy is a relatively good approximation also for the risky portfolios. However, when the credit quality decreases, the importance of the correct exercise policy increases. This is also reflected in Tables \ref{tab_no_net} and \ref{tab_net} in which the corresponding CVA-values are displayed. In practice, a counterparty with a bad credit quality is penalized twice; first by paying a larger CVA (justified), and secondly for paying a CVA which is more overestimated (not justified). The effect is even more significant when we look at expected shortfalls. In Figure \ref{ES-fig}, we see that the ES-CVA, at 97.5\% level, is overestimated by between 19\% and 67\% for portfolio without netting and 27\% to 103\% for the portfolio with netting. Bare in mind that for this particular choice of parameters, the overestimation of the CVAs are only 7.4\% and 11.4\%. All values reported in this the section on risky-valuation are results of only one experiment (no Monte-Carlo averages are used). The reason for this is that we are using many samples, and the variance of the results are therefore low. In addition, we have no reference values to compare with and the main point is that we obtain different values depending on what exercise strategy is used, not high accuracy of the estimated entities.

The WWR-parameter $b$ has a relatively small impact. This is not surprising since a positive $b$ gives WWR for call options and Right Way Risk (RWR) for put options and a the other way around with a negative $b$. Since we have a portfolio consisting of a combination of puts and calls, we have a significant off-setting effect.

\section*{Acknowledgments}
This project is part of the ABC-EU-XVA project and has received funding from the European Unions Horizon 2020 research and innovation programme under the Marie Sk\l dowska--Curie grant agreement No 813261.

\appendix
\section{Neural network details} \label{appendix_A}
In this Appendix, we briefly explain the structure of the neural networks used in this paper. We present pseudo code for the algorithm used to find the exercise strategy for the risk-free portfolio (the extensions to the risky-portfolios are straight-forward). Furthermore, it is described how the time 0 value for a risk-free portfolio is computed.
\subsection{Specification of the neural networks used}\label{NN_spec}
For completeness, we introduce all the trainable parameters that are contained in each of the parameters $\theta_1,\theta_2,\ldots,\theta_{N-1}$, and present the structure of the networks. 

In this section, the following notation is used: 
\begin{itemize}
    \item We denote the dimension of the input layers by $\mathfrak{D}^{\text{input}}\in\N$, and we assume the same input dimension for all $n\in\{1,2,\ldots,N-1\}$ networks. The input is assumed to be the market state $x_n^{\text{train}}\in\R^d$, and hence $\mathfrak{D}^{\text{input}}=d$. However, we can add additional information to the input that is mathematically redundant but helps the training, \textit{e.g.,} the immediate pay-off, to obtain as input\footnote{$\text{concat}$ denotes a concatenation of vectors.} $\text{concat}\left(x_n^\text{train}(m),\,\left(g_1\left(T_n,x_n^\text{train}(m)\right),\ldots,g_J\left(T_n,x_n^\text{train}(m)\right
   )^T\right)\right)^T\in\R^{d+J}$, which would give $\mathfrak{D}^{\text{input}}=d+J$;
    \item For network $n\in\{1,2,\ldots,N-1\}$, we denote the number of layers\footnote{ Input and output layers included.} by $\mathfrak{L}_n\in\N$, and for layer $\ell\in\{1,2,\ldots,\mathfrak{L}_n\}$, the number of nodes by $\mathfrak{N}_{\ell,n}\in\N$. Note that $\mathfrak{N}_{n,1}=\mathfrak{D}^{\text{input}}$;
    \item For network $n\in\{1,2,\ldots,N\}$, and layer $\ell\in\{2,3,\ldots,\mathfrak{L}_n\}$ we denote the weight matrix, acting between layers $\ell-1$ and $\ell$, by $w_{n,\ell}\in\R^{\mathfrak{N}_{n,\ell-1}\times\mathfrak{N}_{n,\ell}}$, and the bias vector by $b_{n,\ell}\in\R^{\ell}$;
    \item For network $n\in\{1,2,\ldots,N\}$, and layer $\ell\in\{2,3,\ldots,\mathfrak{L}_n\}$ we denote the (scalar) activation function by $a_{n,\ell}\colon\R\to\R$ and the vector activation function by $\boldsymbol{a}_{n,\ell}\colon\R^{\mathfrak{N}_{n,\ell}}\to\R^{\mathfrak{N}_{n,\ell}}$, which, for $x=(x_1,x_2,\ldots,x_{\mathfrak{N}_{n,\ell}})$, is defined by 
    \begin{equation*}
        \boldsymbol{a}_{n,\ell}(x)=\begin{pmatrix}a_{n,\ell}(x_1)\\
        \vdots\\
        a_{n,\ell}(x_{\mathfrak{N}_{n,\ell}})\end{pmatrix};
    \end{equation*}
    \item The output of the network should belong to $(0,1)^J\subset\R^J$, meaning that the dimension of the output, denoted by $\mathfrak{D}^{\text{output}}$ should equal $J$. To enforce the output to only take on values in $(0,1)$, we restrict ourselves to scalar-activation functions of the form $a_{n,\mathfrak{L}_{n}}\colon\R\to(0,1)$.
\end{itemize}
Network $n\in\{1,2,\ldots\,N-1\}$ is then defined by \begin{equation}
    \boldsymbol{F}^{\theta_n}(\cdot) = L_{n,\mathfrak{L}_n}\circ L_{n,\mathfrak{L}_n-1}\circ\cdots\circ L_{n,1}(\cdot),\label{F_NN}
\end{equation}
where, for $n\in\{1,2,\ldots,N-1\}$ and for $x\in\R^{\mathfrak{L}_{n,\ell-1}}$, the layers are defined as 
\begin{equation*}
L_{n,\ell}(x)=\begin{cases}x,&\text{for }\ell=1,\\ 
    \boldsymbol{a}_{n,\ell}(w_{n,\ell}^Tx+b_{n,\ell}),&\text{for }\ell\geq 2,\end{cases}
\end{equation*}
with $w_{n,\ell}^T$ the matrix transpose of $w_{n,\ell}$. The trainable parameters of network $n\in\{1,2,\ldots,N-1\}$ are then given by the list\begin{equation*}
    \theta_n=\left\{w_{n,2},b_{n,2},w_{n,3},b_{n,3},\ldots,w_{n,\mathfrak{L}_n}, b_{n,\mathfrak{L}_n}\right\},
\end{equation*}
and as already stated, $\Theta_n=\{\theta_n,\theta_{n+1},\ldots,\theta_{N-1}\}$ and $\Theta=\Theta_1$.

\subsection{Training and valuation}\label{train_NN}
The main idea of the training and valuation procedure is to fit the parameters to some training data, and then use the fitted parameters to make informed decisions with respect to some unseen, valuation data independent of the training data. The training and valuation is described for the risk-free problem, but the procedure is similar for the risky portfolios.
\newline\newline
\noindent\textbf{Training}:\newline
Sample $M_{\text{train}}\in\mathbbm{N}$ independent realizations of $X$, which for $m\in\{1,\,2,\,\ldots,\,M_{\text{train}}\}$ are denoted by $x^{\text{train}}(m)$. In practice, we often need to approximate $X$, \textit{e.g.,} if $X$ satisfies a stochastic differential equation (SDE) we may have to use a temporal discretization scheme. Since this is not the focus of this paper, we assume that $X$ can be approximated pathwise arbitrarily well.

At maturity of the portfolio, we define the cash-flow corresponding to the $j$:th contract and the $m$:th realization of $X$ as $\text{CF}_{N,j}(m)=D_{N-1,N}g_j(T_N,x_N^{\text{train}}(m))$ (recall that $g_j(t_N,\cdot)\equiv 0$ if $T_N$ is larger than the date of maturity for contract $j$). \newline
For $n=N-1,\,N-2,\,\ldots,\,1$, do the following:
\begin{enumerate}
    \item\label{loss_emp}Approximate the optimal parameter $\theta_n^{**}\in\R^{q_n}$, given by
\begin{align*} \theta_n^{**}\in&\argmax_{\theta\in\mathbbm{R}^{q_n}}\bigg(\frac{1}{M_\text{train}}\sum_{m=1}^{M_\text{train}}\sum_{j=1}^JF_j^{\theta}(x_n^\text{train}(m))\,g_j(T_n,x_n^\text{train}(m))\\
&+\big(1-F_j^{\theta}(x_n^\text{train}(m))\big)\text{CF}_{n+1,j}(m)\bigg),\end{align*}
with $\boldsymbol{F}_n^{\theta}$ as in \eqref{F_NN}. The approximation of $\theta_n^{**}$ is denoted by $\theta_n^*$. The above corresponds to an (empirical) loss-function of the form 
\begin{align*}
    L(\theta;x^{\text{train}}) =&- \frac{1}{M_\text{train}}\sum_{m=1}^{M_\text{train}}\sum_{j=1}^JF_j^{\theta}(x_n^\text{train}(m))\,g_j\left(T_n,x_n^\text{train}(m)\right)\\
    &+\big(1-F_j^{\theta}(x_n^\text{train}(m)\big)\text{CF}_{n+1,j}(m).
\end{align*}
In the above we distinguish between the theoretically optimal parameter, $\theta_n^{**}$ (which we rarely find), and the parameter obtained via an optimization algorithm $\theta_n^*$.   
The minus sign in the loss-function transforms the problem from a maximization to a minimization problem, which is the standard formulation in the machine learning community. Note the straightforward relationship between the loss function and the average cash-flows in \eqref{last_step_empirical}. In practice, the data is often divided into mini-batches, for which the loss-function is minimized consecutively. 
\item For $m=1,\,2,\,\ldots,\,M_{\text{train}}$, and for $j=1,2\ldots,J$, update the discounted cash-flows according to: \begin{align*}
    \text{CF}&_{n,j}(m)\\
    =&f_{j}^{\theta_n^*}(x_n^\text{train}(m))\, g(T_n,x_n^\text{train}(m))+D_{n,n+1}\big(1-f_j^{\theta_n^*}(x_n^\text{train}(m))\big)\text{CF}_{n+1,j}(m).
    \end{align*}
\end{enumerate}
Note that we use the continuous versions, $F_j^{\theta_n^*}$, in the optimization phase, and the discontinuous versions $f_j^{\theta_n^*}$ when we update the cash-flows.
The performance of the algorithm does not seem to be sensitive to the specific choice of the number of hidden layers, number of nodes, optimization algorithm, etc. Below is a list of the most relevant parameters/structural choices:
\begin{itemize}
    \item Initialization of the trainable parameters, where a typical procedure is to initialize the biases to 0, and sample the weights independently from a normal distribution; 
    \item The activation functions $a_{\ell,n}$, which are used to add a non-linear structure to the neural networks. In our case we have the strict requirement that the activation function of the output layer maps $\R$ to $(0,1)$. This could, however, be relaxed as long as the activation function is both upper and lower bounded, since we can always scale and shift such output to take on values only in $(0,1)$. For a discussion on different activation functions, see \textit{e.g.,} \cite{activation_functions};
    \item The batch size, $\mathcal{B}_{n}\in\{1,2,\ldots,M_\text{train}\}$, is the number of training samples used for each update of $\theta_n$, \textit{i.e.,} with $\mathcal{B}_n=M_\text{train}$, the loss function is of the form defined in step \ref{loss_emp} above. If we want all batches to be of equal size, we need to choose $\mathcal{B}_n$ to be a multiplier of $M_\text{train}$;
    \item For each update of $\theta_n$, we use an optimization algorithm, for which a common choice is the Adam optimizer, proposed in \cite{Adam}. Depending on the choice of optimization algorithm, there are different parameters related to the specific algorithm to be chosen. One example is the so-called learning rate which decides how much the parameter, $\theta_n$, is adjusted after each batch.
\end{itemize}
Once the parameters have been optimized we define $\Theta^*\coloneqq\{\theta_1^*,\theta_2^*,\ldots,\theta_{N-1}^*\}$, which contains all the information needed in order to use the algorithm for valuation.
\begin{remark}
In the training for the risky portfolio with netting, the algorithm needs to be adjusted since it depends on the exercise history. This cannot easily be included since the algorithm is carried out backwards in time recursively and therefore, at exercise date $T_n$, we do not yet know which derivatives have been exercised prior to $T_n$. This is resolved by, at each exercise date $T_n$, randomly assigning a state for $\alpha_n(m)\in\{0,1\}^J$, representing $A_n$, for each sample $m$. Since the future cash-flows depend on $\alpha_n(m)$, they need to be re-iterated for each $m$. This is done by iteratively, evaluating the already approximated decision functions at $X$ for each exercise date greater than $T_n$, until all the derivatives are exercised, or a default of the counterparty occurs. 
\end{remark}

\noindent\textbf{Valuation}:\newline
Sample $M_\text{test}\in\mathbbm{N}$ independent realizations of $X$, denoted $\left(x_{t}^\text{val}(m)\right)_{t\in[0,T]}$. Denote the vector of decision functions by \begin{equation*}
    \boldsymbol{\mathbbm{f}}^{\Theta^*}_n=\left(\boldsymbol{f}^{\theta_n^*},\,\boldsymbol{f}^{\theta_{n+1}^*},\ldots,\boldsymbol{f}^{\theta_{N-1}^*}\right),\end{equation*}
    and $\boldsymbol{\mathbbm{f}}^{\Theta^*}\coloneqq\boldsymbol{\mathbbm{f}}^{\Theta^*}_1$. We then obtain for sample $m$, \textit{i.e.,} $x^{\text{val}}(m)$, the following stopping rule
\begin{equation*}
    \label{taustar}
    \bar{\tau}_{n,j}^{\Theta^*}=\left(\boldsymbol{\tau}\left[\boldsymbol{\mathbbm{f}}^{\Theta^*}_n\right]\left(x^{\text{val}}(m)\right)\right)_j = \sum_{k=n}^NT_k\,f^{\theta_k^*}_j\big(x^\text{val}_k(m)\big)\prod_{\ell=1}^{k-1}\Big(1-f^{\theta_\ell^*}_j\big(x^\text{val}_\ell(m)\big)\Big).
\end{equation*} The estimated portfolio value at $t=0$ is then given by\begin{equation}\label{pi0_est}
\Pi^\textit{\tiny{V}}(0,x_0) \approx \Upsilon^\textit{\tiny{V}}\big(0,x_0\,|\,\boldsymbol{\mathbbm{f}}^{\Theta^*}\big)\approx \frac{1}{M_\text{val}}\sum_{m=1}^{M_\text{val}}\sum_{j=1}^JD_{0,\bar{\tau}_{0,j}^{\Theta^*}}\,g_j\Big(\bar{\tau}_{0,j}^{\Theta^*},x^\text{val}_{\bar{\tau}_{0,j}^{\Theta^*}}(m)\Big).\end{equation} 
By construction, any stopping strategy is sub-optimal, implying that the estimate \eqref{pi0_est} is biased low. However, it should be pointed out that it is possible to derive a biased-high estimation of $\Pi^\textit{\tiny{V}}(0,x_0)$ from a dual formulation of the optimal stopping problem, which is described in \cite{DOS}. In addition, numerical results in \cite{DOS} show a tight interval for the biased low and biased high estimates for a wide range of problems.

\end{document}